\documentclass{pasa}%
\usepackage{graphicx}
\usepackage{lscape}
\usepackage{caption}
\usepackage{makecell}
\usepackage{placeins}
\usepackage[colorlinks=true,linkcolor=blue,citecolor=blue]{hyperref}

\title[Prediction of Exoplanets]{Exoplanets Prediction in Multiplanetary Systems}

\author[Mousavi-Sadr, Gozaliasl, and Jassur]{M.~Mousavi-Sadr$^{1,3}$, G.~Gozaliasl$^2$ and D.~M.~Jassur$^1$
\affil{$^1$Department of Theoretical Physics and Astrophysics, Faculty of Physics, University of Tabriz, Tabriz, Iran}%
\affil{$^2$Department of Physics, University of Helsinki, P. O. Box 64, FI-00014, Helsinki, Finland}
\affil{$^3$Email: mahdiyar.mousavi@tabrizu.ac.ir}
}

\jid{PASA}
\doi{10.1017/pas.\the\year.xxx}
\jyear{\the\year}

\begin{document}

\defcitealias{2013MNRAS.435.1126B}{BL13}
\defcitealias{2015MNRAS.448.3608B}{BL15}
\defcitealias{2017A&A...604A..83B}{B17}

\begin{frontmatter}
\maketitle

\begin{abstract}
We present the results of a search for additional exoplanets in  all  multiplanetary systems discovered to date, employing a logarithmic spacing between planets in our Solar System known as the Titius-Bode (TB) relation.  We use the Markov Chain Monte Carlo method and separately analyse 229  multiplanetary systems that house at least three or more confirmed planets. We find that the planets in $\sim53\%$ of these systems adhere to a logarithmic spacing relation remarkably better than the Solar System planets. Using the TB relation, we predict the presence of 426 additional exoplanets in 229 multiplanetary systems, of which 197 candidates are discovered by interpolation and 229 by extrapolation. Altogether, 47 predicted planets are located within the habitable zone (HZ) of their host stars, and five of the 47 planets have a maximum mass limit of 0.1-2$M_{\oplus}$ and a maximum radius lower than 1.25$R_{\oplus}$. Our results and prediction of additional planets agree with previous studies' predictions; however, we improve the uncertainties in the orbital period measurement for the predicted planets significantly.
\end{abstract}

\end{frontmatter}

\section{INTRODUCTION}
The number of detected exoplanets is growing rapidly, so that over four thousand exoplanets have been detected and confirmed to date. There are also thousands of other candidate exoplanets that require further follow-up observation. The Kepler spacecraft plays a key role in detecting these systems. The Kepler space mission's main goal was to discover Earth-size exoplanets in or near the habitable zone (HZ) of solar-like stars and determine the fraction of the hundreds of billions of stars in our galaxy that might possess such planets \citep{1998SPIE.3356..599K}.

Photometry with the transit method is the most successful exoplanet discovery method, which has been used by Kepler space mission, Transiting Exoplanet Survey Satellite (TESS), and many ground-based observatories \citep{2010Sci...327..977B,2015JATIS...1a4003R,2018haex.bookE.117D}. However, this method has its own difficulties. For example, transits are detectable only when the planet's orbit happens to be almost exactly aligned with the observer's line-of-sight. This covers only a small fraction of exoplanets. Furthermore, the planet's transit lasts for a small fraction of its total orbital period. As a result, it is not very likely to detect planets' transits, especially those with long orbital periods. This study sets out to predict the existence of additional undetected planets in multiple exoplanet systems.

In our Solar System, there is a simple logarithmic spacing between planets, which has been known for over two centuries as the Titius-Bode (TB) law. Its classical relation is:
\begin{equation}
a_{n}=0.4+0.3\times2^{n},
\end{equation}
where $a_{n}$ represents the semi-major axis of the $n^{th}$ planet in AU. The planet Mercury corresponds to $n=-\infty$, Venus to $n=0$, Earth to $n=1$ and so on \citep{1974PhT....27e..54N}. After the discovery of the planet Uranus in 1781 by Frederick William Herschel, it was recognised that the TB law predicted this planet's semi-major axis \citep{1974PhT....27e..54N}.

The discovery of the TB relation motivated many observation programs to investigate and detect the lost fifth planet, which eventually led to the discovery of the asteroid Ceres \citep{1948JRASC..42..241S}. The predictions made using the TB relation also played a key role in exploring the planet Neptune, but not as accurately as Uranus. Interestingly, the satellite systems of the giant planets also follow a TB relation \citep{1960VA......3...25L, 1970CeMec...3...67B}.

\par The TB relation was used effectively to predict lost undetected objects in our Solar System. It was believed that this relation could help make similar predictions in detected multiple exoplanet systems, too. The five-planet 55 Cnc system was one of the first multiple- exoplanet systems where \citet{2008RMxAA..44..243P} applied TB to predict the undetected planets. They found that a simple exponential TB relation reproduces the semi-major axes of the five observed planets. They also predicted two additional planets at distances of 2 and 15 AU. Using the 55 Cnc system, \citet{2008JASS...25..239C} also checked whether the TB relation is enforceable on exoplanetary systems by statistically analysing the distribution of the ratio of periods of two planets in the 55 Cnc system, by comparing it with that derived from the TB relation. \citet{2010JASS...27....1C} again repeated this calculation for 31 multiple exoplanet systems and concluded that the adherence of the Solar System's planets to the TB relation might not be fortuitous; thus, we could not ignore the possibility of using the TB relationship in exoplanetary systems.

Moreover, \citet{2012AIPC.1479.2356L} showed that like 55 Cnc, ten other planetary systems (ups And, GJ 876, HD 160691, GJ 581, Kepler-223, HR 8799, Kepler-20, Kepler-33, HD 10180, and Kepler-11) host four or more planets that also obey a similar (but not identical) TB relation.

\citet{2013MNRAS.435.1126B} (hereafter, \citetalias{2013MNRAS.435.1126B}) used a sample of 68 multiple exoplanet systems with at least four planets, including samples of both confirmed and candidate systems. They identified a sample of exoplanet systems that are likely to be more complete and tested their adherence to the TB relation. They found that most of these exoplanetary systems adhere to the TB relation better than the Solar System. Using a generalized TB relation, they predicted 141 additional exoplanets, including a planet with a low radius ($R<2.5R_{\oplus}$) and within the HZ of the Kepler-235.

Using the predictions made by \citetalias{2013MNRAS.435.1126B}, \citet{2014MNRAS.442..674H} analysed Kepler's long-cadence data to search for 97 of those predicted planets and obtained a detection rate of $\sim5$ percent. Considering the possibility that the remaining predicted planets were not detected because of their small size or non-coplanarity, the detection rate was estimated to be less than the lower limit of the expected number of detections. They concluded that applying the TB relation to exoplanetary systems and using its predictive power in \citetalias{2013MNRAS.435.1126B} could be questionable.

\begin{figure}
 \includegraphics[width=\columnwidth]{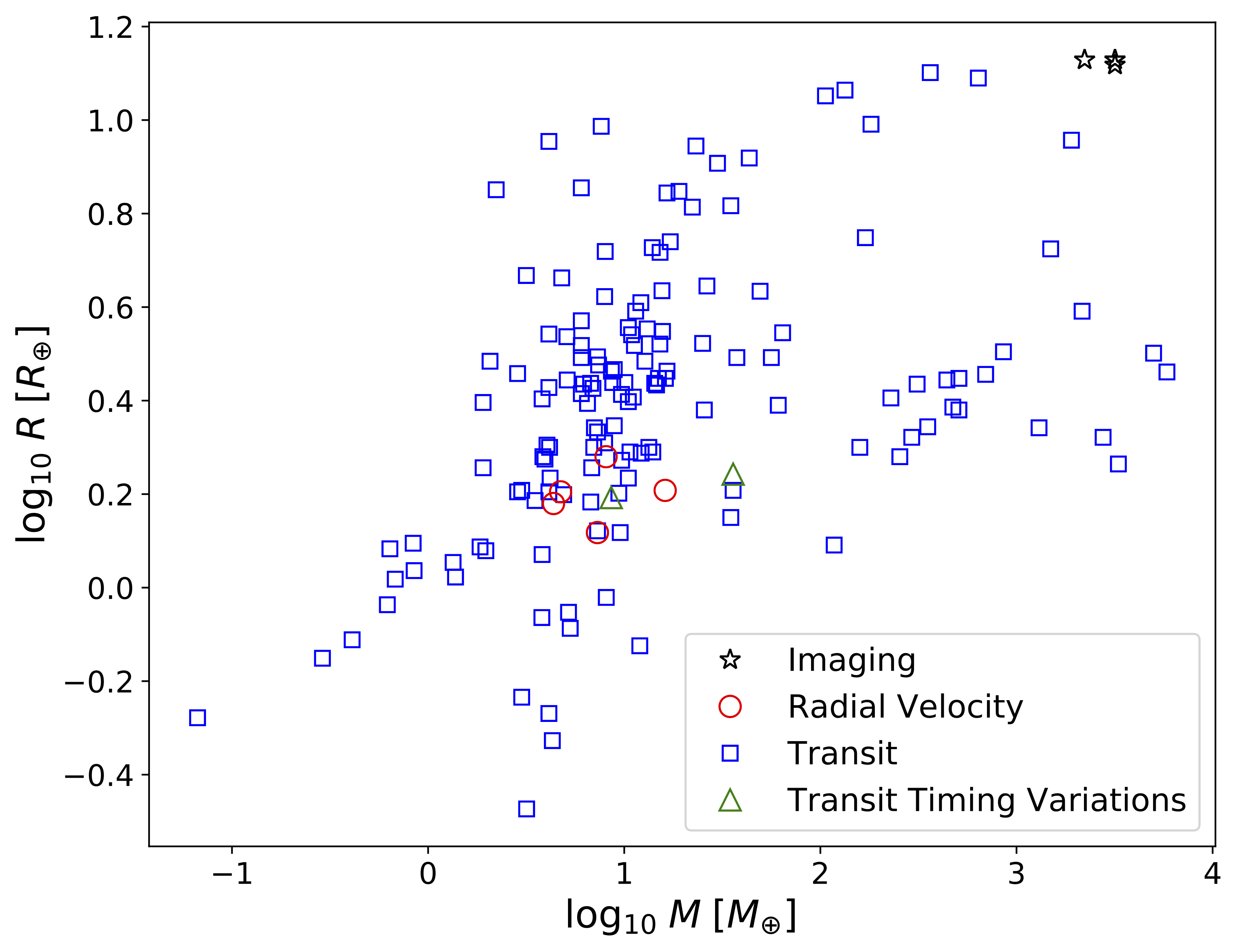}
 \caption{The mass-radius distribution of the exoplanets in our sample, separated into five groups based on their detection methods: imaging (black stars), radial velocity (red circles), transit (blue squares), and transit timing variations (green triangles).}
 \label{figure1}
\end{figure}

\citet{2015MNRAS.448.3608B} (hereafter, \citetalias{2015MNRAS.448.3608B}) used the \citetalias{2013MNRAS.435.1126B} method to predict the existence of additional planets in 151 Kepler multiple exoplanet systems that contain at least three transiting planets. They found 228 undetected exoplanets and, on average, $2\pm1$ planets in the HZ of each star. They found that apart from the five planets detected by \citet{2014MNRAS.442..674H}, one additional planet in Kepler-271 had been detected around its predicted period. The completeness of observational data was taken into account as an important issue in their analysis. They declared a completeness factor due to the intrinsic noise of the host stars, planet radius, highly inclined orbits, and the detection techniques. Thus they did not expect all predicted planets to be detected. They estimated the geometric probability to transit for all 228 predicted planets and highlighted a list of 77 planets with high transit probability, resulting in an expected detection rate of $\sim15$ percent, which was about three times higher than the detection rate measured \citetalias{2013MNRAS.435.1126B}.

Recently, \citet{2020PASJ...72...24L} used data from 27 exoplanetary systems with at least five planets and applied their proposed method to find the reliability of the TB relation and its predictive capability to search for planets. They removed planets from the system one by one and used TB relation to recover them, where they were able to recover the missing one 78\% of the time. This number was much higher than when they tried this with random planetary systems, where 26\% of planets were recovered. Using statistical tests, they showed that the planetary orbital periods in exoplanetary systems were not consistent with a random distribution and concluded it to be an outcome of the interactions between true planets.

The purpose of this study is to test the adherence of 229 multiple exoplanet systems with at least three detected planets to the TB relation, and compare their adherence rate with the Solar System's using Markov chain Monte Carlo (MCMC) \citep[see][]{goodman10,emcee3} as a precise method for regression while considering the reported uncertainties of physical values. We also aim to predict the existence of additional undetected exoplanets and estimate their physical properties, either maximum mass or maximum radius. Moreover, we highlight exoplanets located within the HZ of their host stars. Using a sample of seven multiple exoplanet systems with detected planets after predictions made by \citetalias{2015MNRAS.448.3608B}, we also aim to determine if the TB relation is a reliable method of identifying undetected member planets. Finally, we compare our predictions with those from \citetalias{2015MNRAS.448.3608B}.

This paper is organised as follows: Section \ref{data} describes the data and methods used to predict exoplanets and estimate their maximum mass and maximum radius. Section \ref{results} presents our results. Section \ref{summary} summarises our main results and conclusions.

\section{DATA AND METHOD}\label{data}
\subsection{Data}
We used physical parameters of exoplanets, including planetary orbital period, radius, and mass, and the stellar mass and radius from two exoplanet databases: the NASA Exoplanet Archive\footnote{\url{http://exoplanetarchive.ipac.caltech.edu/}} and the Extrasolar Planets Encyclopedia\footnote{\url{http://exoplanet.eu/}}. We also used conservative and optimistic limits of the HZ of available stars from the Habitable Zone Gallery\footnote{\url{http://hzgallery.org/}}. 

\begin{figure}
 \includegraphics[width=\columnwidth]{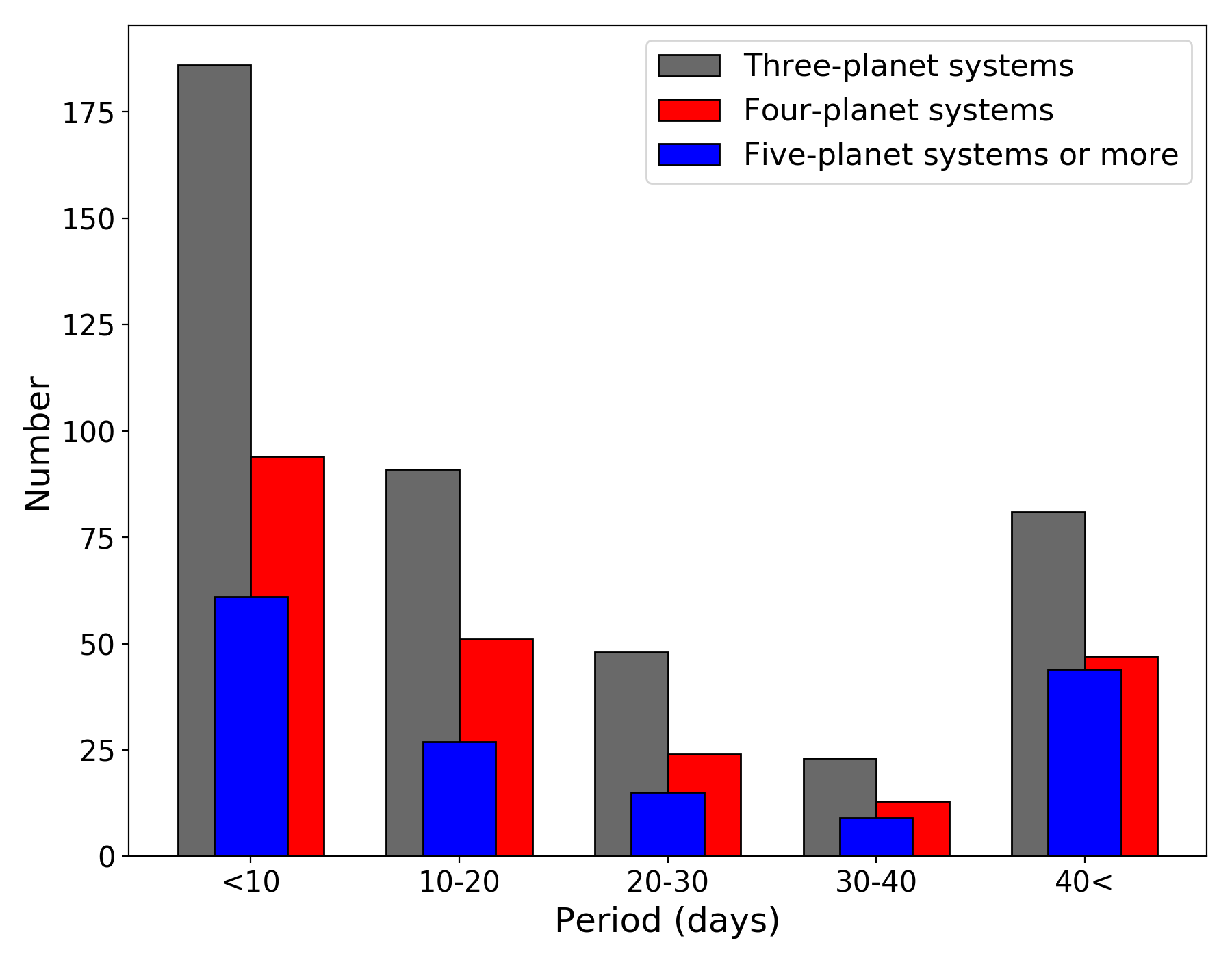}
 \caption{The distribution of the orbital period of 813 member exoplanets in 229 multiple-exoplanet systems hosting at least three planets (gray bars), four planets (red bars), and five (or more) planets (blue bars).}
 \label{figure2}
\end{figure}

The total number of multiple exoplanet systems with at least three confirmed planets available is 230 systems to date, hosting 818 planets. 81.5\% of these planets have been detected using the transit method and 16.4\% using the radial velocity method. The remaining 2.1\% of planets have been identified using transit timing variations, imaging, pulsar timing, and orbital brightness modulation methods. Figure \ref{figure1} represents the mass-radius distribution of exoplanets for five groups of planets, separated according to their detection methods. Of the 230 systems, 142 systems host three planets, 59 systems host four planets, 20 systems include five planets, and seven systems host six planets. TRAPPIST-1 is the only system that contains seven planets, and KOI-351 the only one that contains eight. Figure \ref{figure2} represents the distribution of the orbital period of member exoplanets in various exoplanetary systems used in this study. 

The exoplanetary systems 55 Cnc, GJ 676 A, HD 125612, K2-136, Kepler-132, Kepler-296, Kepler-47, Kepler-68, and ups And consist of binary stars. GJ 667 C and Kepler-444 are also triple star systems. The system Kepler-132 possesses four planets such that two of them (Kepler-132 b and Kepler-132 c) have roughly similar orbital periods of 6.1782 days and 6.4149 days. After more detailed studies, it was found that Kepler-132 b and Kepler-132 c cannot orbit the same star \citep{2014ApJ...784...44L}. Consequently, we exclude Kepler-132 from our analyses; our final sample of exoplanetary systems thus contains 229 systems.

\begin{figure}
 \includegraphics[width=\columnwidth]{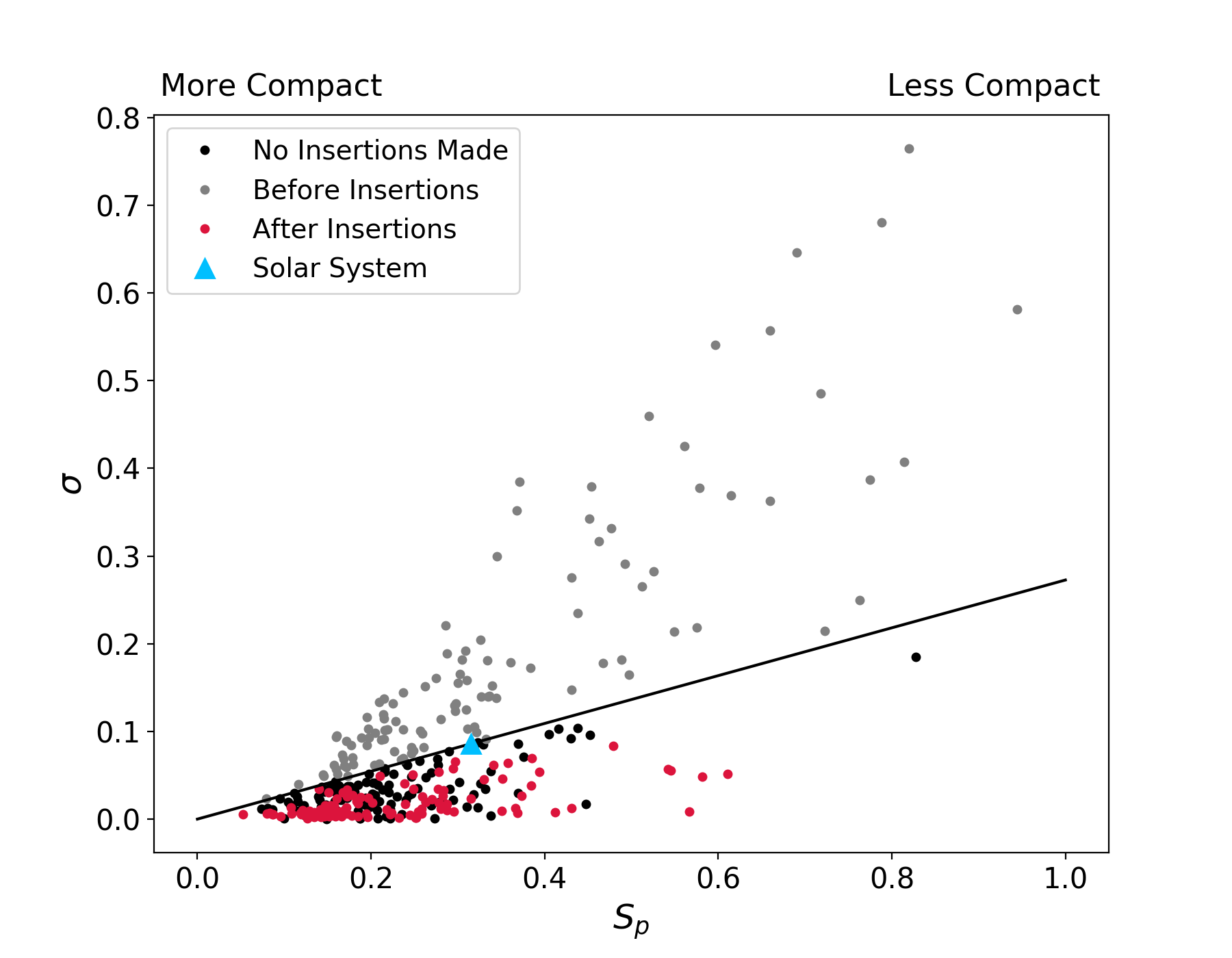}
 \caption{$\sigma$ as a function of the average log period spacing between planets, $S_p$ (Equation \ref{equation5}), of exoplanet systems. The black line goes through two points: the origin (0,0) and ($S_{p}$,$\sigma$), where $S_{p}$ is the compactness of the Solar System and $\sigma$ is the value required for the Solar System to yield $\chi^2/dof=1$ in Equation \ref{equation4}. The cyan triangle shows the Solar System, and the black dots show the exoplanet systems with no planet insertions (systems with $\chi^2/dof\leq1$). Gray dots indicate systems before planet insertions, and red dots indicate the ($S_{p}$,$\sigma$) of the systems after insertions have been made.}
 \label{figure3}
\end{figure}

\begin{figure*}
\centering
 \includegraphics[width=0.55\paperwidth]{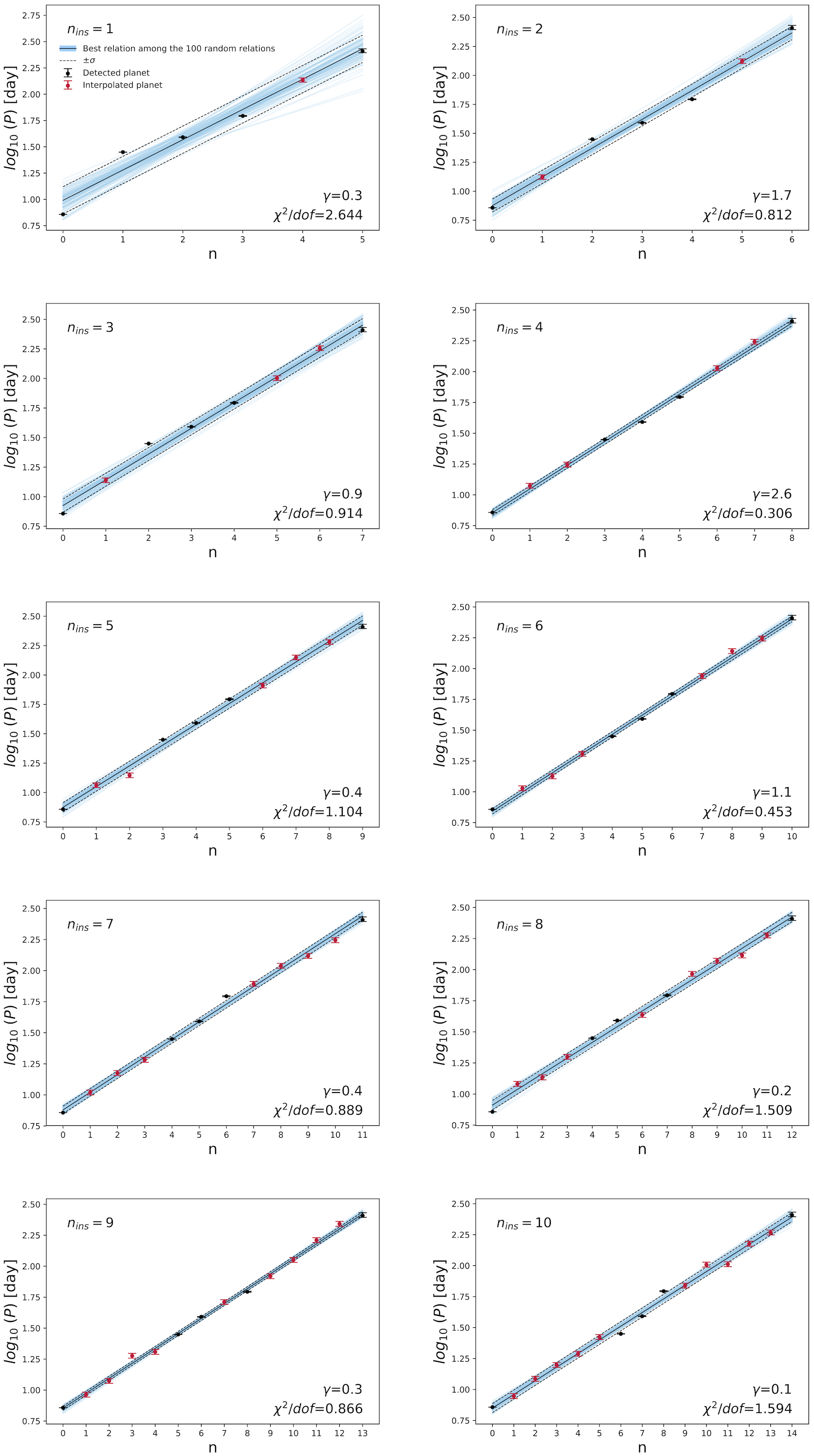}
 \caption{The TB relation and steps of linear regressions are applied to the data of system GJ 667 C . Detected and predicted planets are shown with black and red dots, respectively. For each step, the value of $n_{ins}$ represents the number of inserted planets. The values of $\gamma$ and $\chi^2/dof$ are also shown where the highest value of $\gamma$ is in the fourth step as the best combination of detected and predicted planets. The two black dashed lines show $\pm1\sigma$ uncertainties around the best scaling relation (black solid line). The blue lines are a set of 100 different realisations, drawn from the multivariate Gaussian distribution of the parameters (for the fourth step: m=0.1924, b=0.856 and ln<$\sigma$>=-3.73), and the scatter co-variance matrix is estimated from the MCMC chain.}
 \label{figure4}
\end{figure*}

\subsection{Methods}
\subsubsection{The TB relation and the prediction of additional exoplanets in multiple-exoplanet systems}
We use the TB relation to predict additional undetected planets for all multiple exoplanet systems, with at least three confirmed planets.
\par The TB relation can be written in terms of the orbital periods as follows:
\begin{equation}
P_{n}=P\alpha^{n},   n=0,1,2,...,N-1.
\end{equation}
Where $P_{n}$ is the orbital period of the $n^{th}$ planet, and P and $\alpha$ are fitting parameters. In analysing the exoplanetary systems, the Solar System is used to guide how well each system adheres to the TB relation. In logarithmic space, the TB relation is written as follows:
\begin{equation}
    \log P_{n}=\log P + n\log\alpha=b+m\times n,   n=0,1,2,...,N-1.
	\label{eq:quadratic}
\end{equation}
Where b=$\log P$ and m=$\log\alpha$ are intercept and slope of the relation, respectively. For a system with N planets, the $\chi^2/dof$ value can be calculated by:
\begin{equation}
\frac{\chi^2(b,m)}{N-2}=\frac{1}{N-2}\sum_{n=0}^{N-1}[\frac{(b+m\times n)-\log P_{n}}{\sigma}]^2,
\label{equation4}
\end{equation}
N-2 is the number of degrees of freedom (N planets - 2 fitted parameters (b and m)), and $\sigma$ represents the system's sparseness or compactness. The sparseness/compactness of a system is calculated from $\sigma$= 0.273 $S_p$, where $S_p$ represents the average log period spacing between planets as defined by:
\begin{equation}
S_{p}=\frac{\log P_{N-1}-\log P_{0}}{N},
\label{equation5}
\end{equation}
where N is the number of planets in the system and $P_{N-1}$, and $P_{0}$ are the largest and smallest orbital periods in the system, respectively. We plot the $\sigma$ values (see Equation \ref{equation4}) as a function of sparseness/compactness ( $S_{p}$; see Equation \ref{equation5}) in Fig. \ref{figure3}. The black line goes through two points: the origin and the specific $S_{p}$ and $\sigma$ values for the Solar System ($\sigma$ is the value required for Solar System to yield $\chi^2/dof=1$ in Equation \ref{equation4}). The black points show the exoplanet systems with no planet insertions, gray points indicate systems before planet insertion, and red points indicate systems after insertions have been made.

Using a proper value for $\sigma$, $\chi^2/dof=1$ is adjusted for the Solar System case. If the detected planets in a system adhere to the TB relation better than the planets of the Solar System ($\chi^2/dof\leq1$), we only predict an extrapolated planet beyond the outermost detected planet. If the detected planets adhere worse ($\chi^2/dof>1$), we begin the interpolation process; for the first step of interpolation, one new specific planet, from at-least 5,000 hypothetical planets that have a random period between the innermost and outermost detected planets, is inserted into the system covering all possible locations between the two adjacent planets, and a new $\chi^2/dof$ value calculated for each possibility. The new specific inserted planet is chosen from 5,000 cases when producing the minimum value of $\chi^2/dof$. Similarly, inserting up to 10 new specific planets for each system step by step (two planets for the second step, three planets for the third step, etc.) covers all possible locations and combinations between two adjacent planets. The period uncertainty of an inserted planet ($e_{ins}$) is calculated using the uncertainties of detected planets in the same system (e.g., $e_{1}$, $e_{2}$, $e_{3}$,...) as follows:
\begin{equation}
e_{ins}=\sqrt{e_{1}^{2}+e_{2}^{2}+e_{2}^{3}+...}.
\end{equation}
We adopt the highest value of the parameter $\gamma$, which is the improvement in the $\chi^2/dof$ per inserted planet, for identifying the best combination of detected and predicted planets. $\gamma$ is defined by:
\begin{equation}
\gamma=\frac{(\frac{\chi_{i}^2-\chi_{j}^2}{\chi_{j}^2})}{n_{ins}},
\end{equation}
where $\chi_{i}^2$ and $\chi_{j}^2$ are the $\chi^2$ values before and after inserting of $n_{ins}$ planets.

We analyse each system's data separately and use the MCMC method to quantify the uncertainties of the best-fit parameters.

\begin{table}
\small
\caption{Data corresponding to Fig. \ref{figure4}.}
\label{table1}
\centering
\begin{tabular}{lllll}
\hline \hline
$n_{ins}$$^a$ & $\chi^2/dof$ & $\gamma$$^{b}$ & Period\;(days) & ON$^c$\\
\hline
1     & 2.644 & 0.3   & 136.5 & 4 \\
2     & 0.812 & 1.7   & 13.3  & 1 \\
          &       &       & 132.6 & 5 \\
3     & 0.914 & 0.9   & 13.8  & 1 \\
          &       &       & 100.6 & 5 \\
          &       &       & 180.4 & 6 \\
4     & 0.306 & 2.6   & 11.9  & 1 \\
          &       &       & 17.6  & 2 \\
          &       &       & 106.7 & 6 \\
          &       &       & 174.8 & 7 \\
5     & 1.104 & 0.4   & 11.5  & 1 \\
          &       &       & 14.0    & 2 \\
          &       &       & 81.4  & 6 \\
          &       &       & 141.0   & 7 \\
          &       &       & 190.7 & 8 \\
6     & 0.453 & 1.1   & 10.7  & 1 \\
          &       &       & 13.4  & 2 \\
          &       &       & 20.3  & 3 \\
          &       &       & 86.9  & 7 \\
          &       &       & 138.6 & 8 \\
          &       &       & 175.1 & 9 \\
7     & 0.889 & 0.4   & 10.5  & 1 \\
          &       &       & 15.0    & 2 \\
          &       &       & 19.1  & 3 \\
          &       &       & 78.3  & 7 \\
          &       &       & 109.3 & 8 \\
          &       &       & 131.3 & 9 \\
          &       &       & 175.6 & 10 \\
8     & 1.509 & 0.2   & 12.1  & 1 \\
          &       &       & 13.6  & 2 \\
          &       &       & 19.9  & 3 \\
          &       &       & 43.2  & 6 \\
          &       &       & 92.5  & 8 \\
          &       &       & 117.9 & 9 \\
          &       &       & 130.0   & 10 \\
          &       &       & 188.5 & 11 \\
9     & 0.866 & 0.3   & 9.2   & 1 \\
          &       &       & 11.8  & 2 \\
          &       &       & 18.9  & 3 \\
          &       &       & 20.3  & 4 \\
          &       &       & 51.4  & 7 \\
          &       &       & 83.0    & 9 \\
          &       &       & 112.3 & 10 \\
          &       &       & 162.5 & 11 \\
          &       &       & 220.0   & 12 \\
10    & 1.594 & 0.1   & 8.8   & 1 \\
          &       &       & 12.2  & 2 \\
          &       &       & 15.8  & 3 \\
          &       &       & 19.4  & 4 \\
          &       &       & 26.5  & 5 \\
          &       &       & 69.0    & 9 \\
          &       &       & 101.9 & 10 \\
          &       &       & 102.6 & 11 \\
          &       &       & 150.3 & 12 \\
          &       &       & 185.2 & 13 \\
\hline \hline
\end{tabular}
\medskip
  \tabnote{$^a$Number of the inserted planet.}
  \tabnote{$^b$$\gamma=(\chi_{i}^2-\chi_{j}^2)/(\chi_{j}^2\times n_{ins})$, where $\chi_{i}^2$ and $\chi_{j}^2$ are the $\chi^2$ values before and after inserting of $n_{ins}$ planets, respectively.}
  \tabnote{$^c$The orbital number of the inserted planet.}
\end{table}

By applying the lowest signal-to-noise ratio (SNR) of the detected planets in the same system to the predicted planet's orbital period, the maximum mass or maximum radius of the predicted planets is calculated.

For transiting detected planets, the maximum radius is calculated by:
\begin{equation}
R_{max}=R_{min SNR}(\frac{P_{predicted}}{P_{min SNR}})^{0.25},
\end{equation}
and for radial velocity detected planets, the maximum mass is calculated by:
\begin{equation}
M_{max}=M_{minSNR}(\frac{P_{predicted}}{P_{min SNR}})^{\frac{7}{6}},
\end{equation}
where $R_{minSNR}$, $M_{minSNR}$, and $P_{minSNR}$ are the radius, mass, and orbital period of the detected planet with the lowest SNR, respectively. After calculating the maximum radius or maximum mass of the predicted planets, using the mass-radius relationship established by \citet{2017A&A...604A..83B} (hereafter, \citetalias{2017A&A...604A..83B}) ($R_{p} \propto M^{0.55}$ and $R_{p} \propto M^{0.01}$ for the small and large planets, respectively), we convert the radius values to mass values, and vice versa.

\subsubsection{Dynamical stability and the transit probability of the predicted planets}
We use the dynamical spacing criterion ($\Delta$) to analyse how our predicted objects could be stable in their positions in the exoplanetary system. The dynamical spacing $\Delta$ between two adjacent planets with masses $M_{1}$ and $M_{2}$ and orbital periods $P_{1}$ and $P_{2}$ orbiting a host star with a mass of $M_{*}$ was defined by \citet{1993Icar..106..247G} and \citet{1996Icar..119..261C} as follows:
\begin{equation}
\Delta=\frac{2M_{*}^{1/3}(P_{2}^{2/3}-P_{1}^{2/3})}{(M_{1}+M_{2})^{1/3}(P_{2}^{2/3}+P_{1}^{2/3})}.
\end{equation}
In a planetary system, two adjacent planets are less likely to be stable in their positions when the value of their dynamical spacing is small ($\Delta \leq 10$). However, if they are stable, they are more likely to be in orbital resonance with each other. Using this criterion, we analyse 45 systems in our sample, which have predicted planets within their HZs. We calculate the $\Delta$ value for all pairs in these systems and investigate whether they are in resonance with each other or not (for planets whose mass values have not been reported in catalogues, we use the mass-radius relation of \citetalias{2017A&A...604A..83B} to calculate their masses). To analyse the possibility of resonance, we calculate the period ratios of all adjacent planets considering an arbitrary threshold; the planet pairs that have $x\leq2\%$ are considered to be pairs in orbital resonance, where
\begin{equation}
x=\frac{\mid \frac{N_{j}}{N_{i}}-\frac{P_{n+1}}{P_{n}}\mid}{\frac{N_{j}}{N_{i}}},
\end{equation}
$N_{i}$ and $N_{j}$ are positive integers with $N_{i}<N_{j}\leq5$, and $P_{n}$ and $P_{n+1}$ are the orbital periods of two adjacent planets.

\begin{figure}
 \includegraphics[width=\columnwidth]{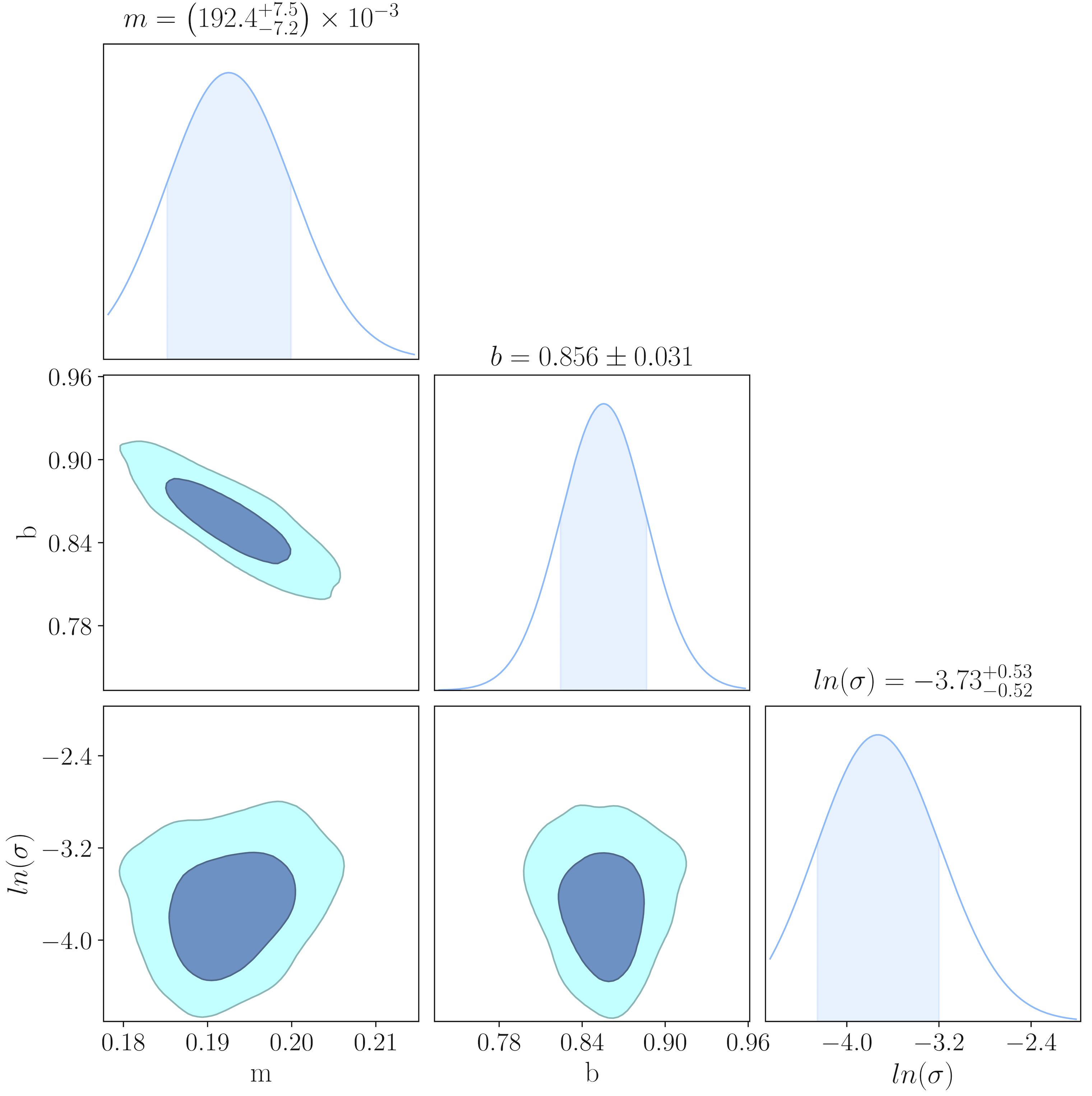}
 \caption{The one- and two-dimensional marginalized posterior distributions of the scaling relation parameters for the highest $\gamma$ value corresponding to the fourth step (($n_{ins}=4$)) of the linear regression of the GJ 667  C system, as shown in Fig. \ref{figure4}.}
 \label{figure5}
\end{figure}

\begin{figure}
 \includegraphics[width=\columnwidth]{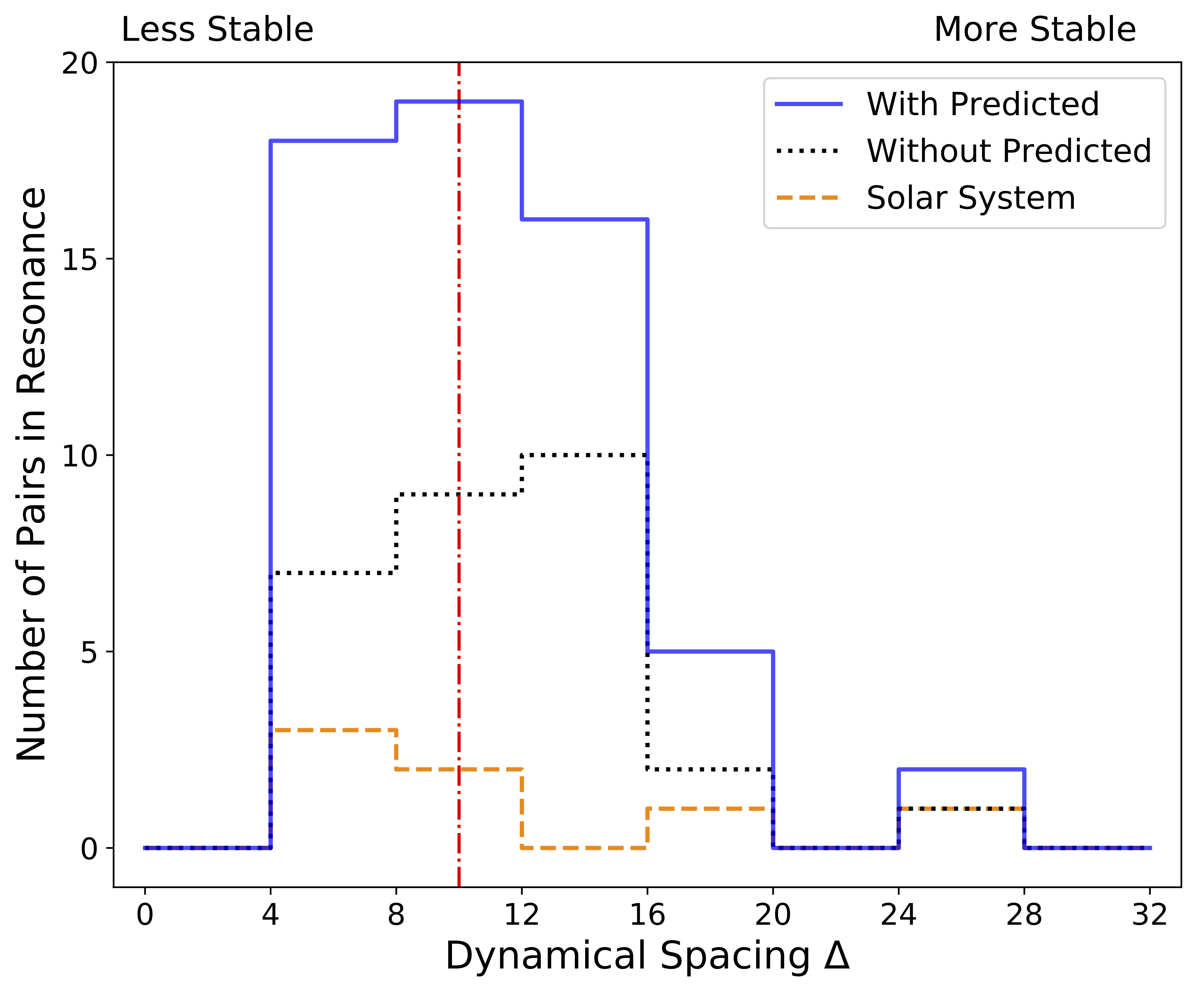}
 \caption{Dynamical spacing $\Delta$ and the total number of adjacent exoplanet pairs that are in orbital resonance with each other. The solid blue line shows the number of resonance pairs considering our predicted planets, and the dotted black line shows the number before inserting any predicted planets into systems. The values for the Solar System are also shown for reference via the orange dashed line. The vertical dash-dotted red line corresponds to $\Delta=10$ and separates the less and more stable adjacent planet pairs regimes.}
 \label{figure6}
\end{figure}

The transit phenomenon of a planet is only observable if the planetary orbit plane is close to the line-of-sight between the observer and the host star. In other words, the pole of the planetary orbit must be within the angle $d_{s}/a_{p}$, where $d_{s}$ is the stellar diameter, and $a_{p}$ is the planetary orbital radius. Then, the geometric transit probability ($P_{tr}$) of a planet can be estimated using $d_{s}/2a_{p}$, where $P_{tr}=0.5\%$ for an Earth-size planet at 1 AU orbiting a Solar-size star \citep{1984Icar...58..121B,1996chz..conf..229K}. According to the models of planetary systems, multi-planetary systems, like the Solar System, are assumed to be formed out of common protoplanetary disks \citep{2015ARA&A..53..409W}. Therefore, the orbital planes should have small relative inclinations so that the Kepler multi-planetary systems are highly coplanar \citepalias{2015MNRAS.448.3608B}. We use this criterion to estimate the transit probability of predicted exoplanets and prioritise how soon the predicted planets can be detected.

\begin{figure*}
\centering
 \includegraphics[width=0.56\paperwidth]{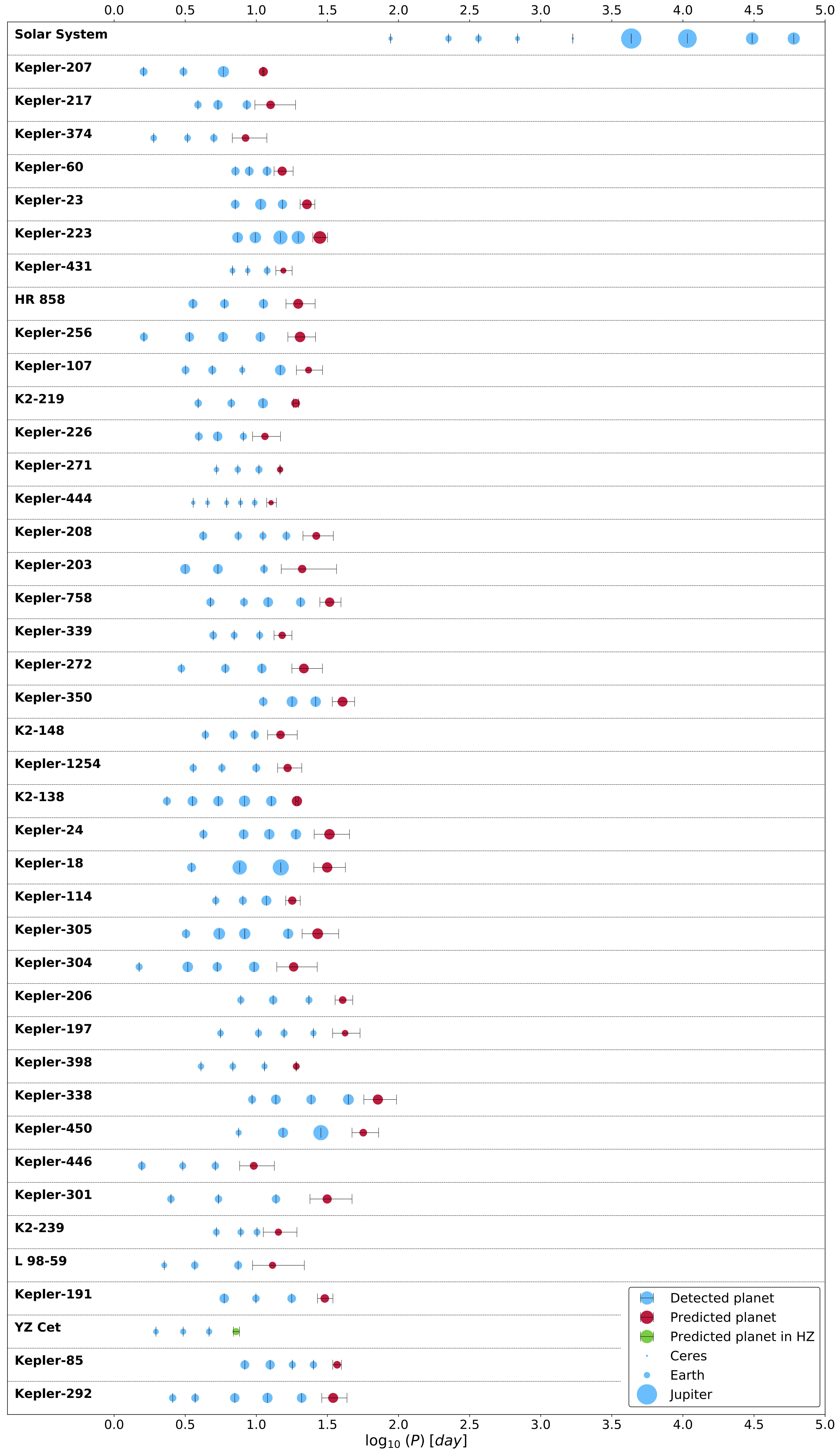}
 \caption{Orbital periods and scaled radii of exoplanets in multiple-planet systems including the predicted exoplanets from extrapolations. The cyan and red circles indicate detected and predicted planets in systems, respectively. The green circles also indicate the predicted planets located within the HZ of their parent stars. The estimated radius of the predicted planet in GJ 676 A is higher than the maximum possible limit of a typical planet. Furthermore, due to the discovery method of HR 8799, the predicted planet's radius is not calculated. Therefore, GJ 676 A and HR 8799 are excluded.}
 \label{figure7}
\end{figure*}

\begin{figure*}
\ContinuedFloat
\begin{center}
 \includegraphics[width=0.56\paperwidth]{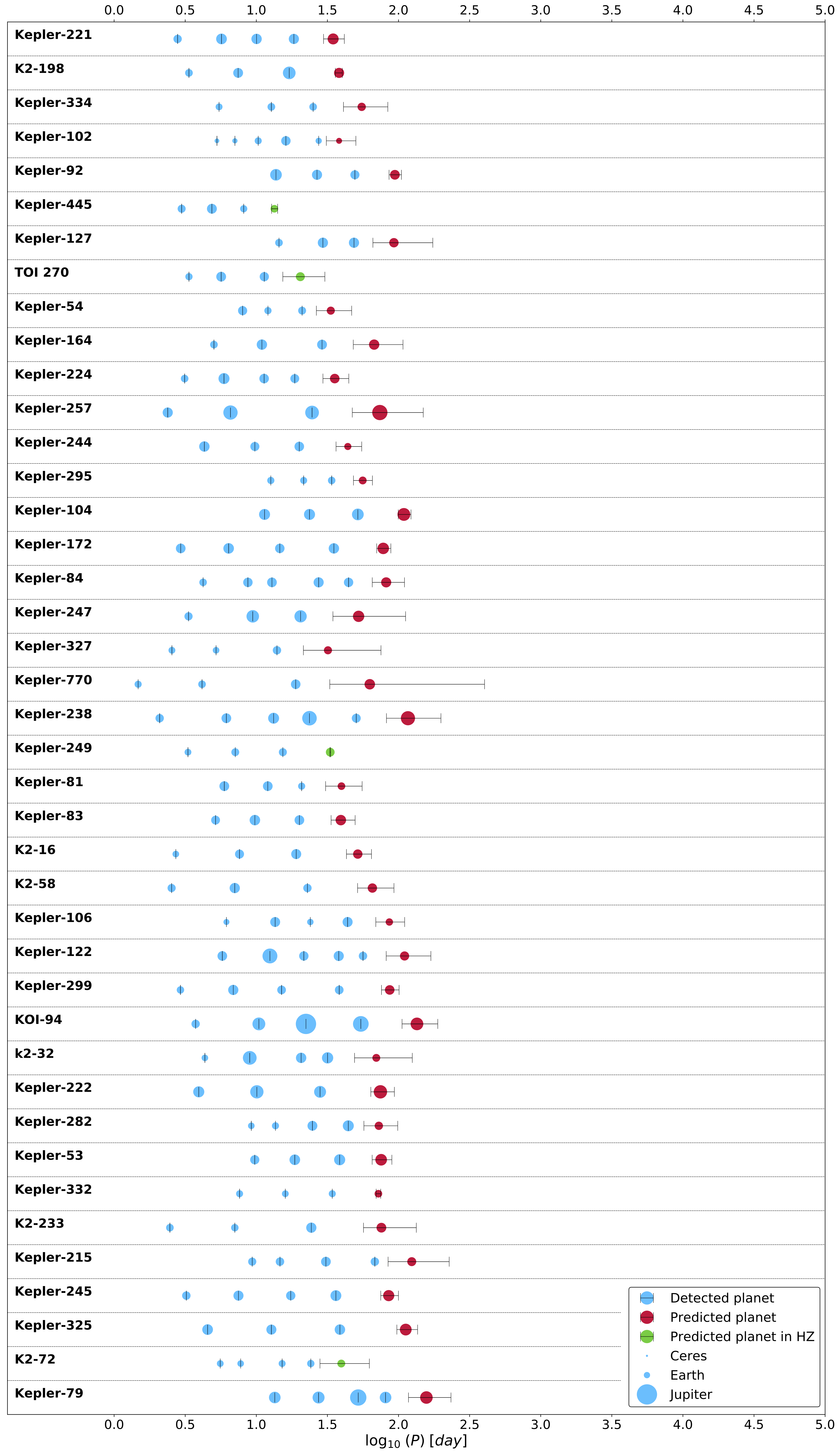}
 \end{center}
 \caption[]{continued}
 \label{}
\end{figure*}

\begin{figure*}
\ContinuedFloat
\begin{center}
 \includegraphics[width=0.56\paperwidth]{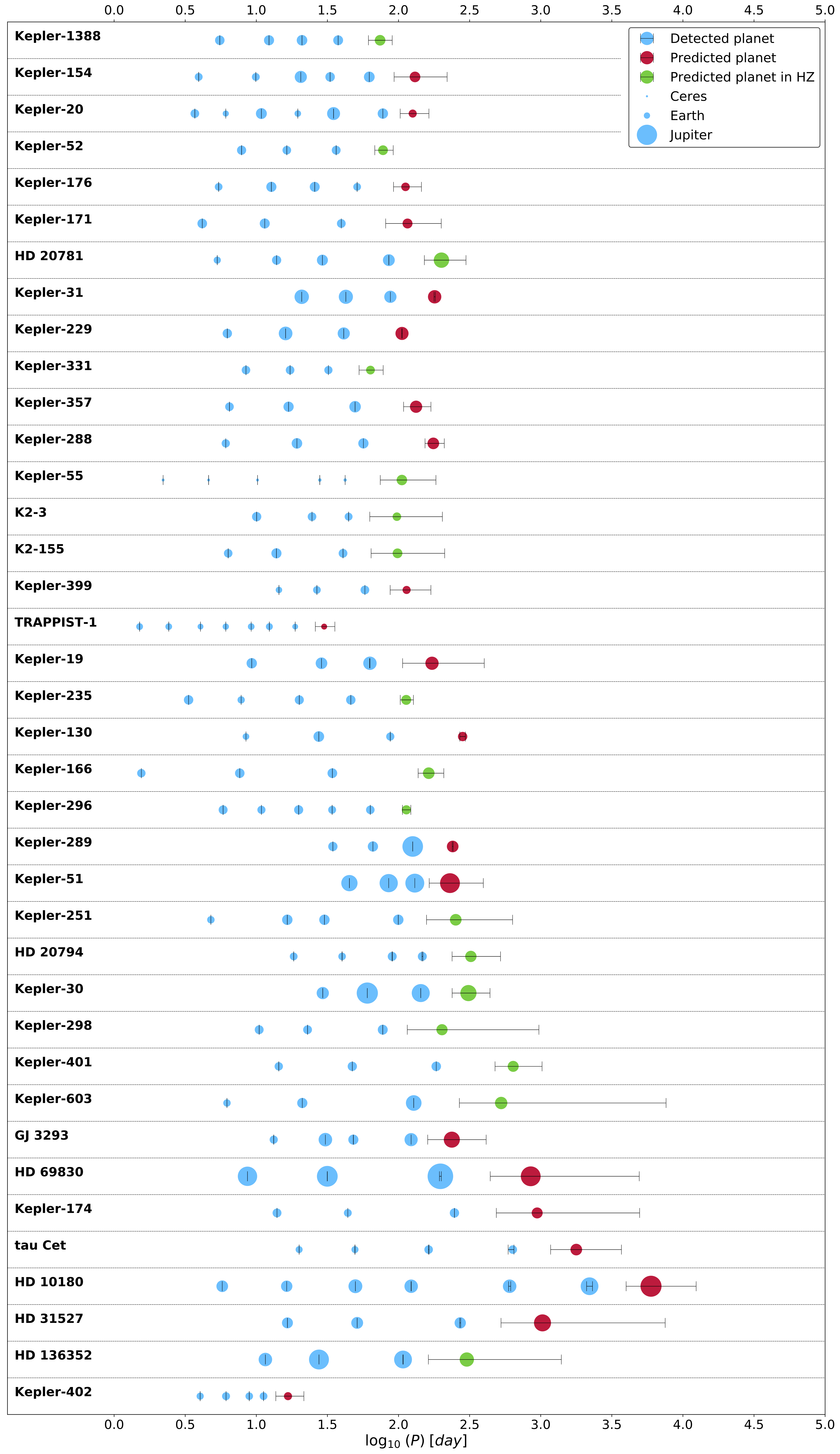}
 \end{center}
 \caption[]{continued}
 \label{}
\end{figure*}

\begin{figure*}
\centering
 \includegraphics[width=0.56\paperwidth]{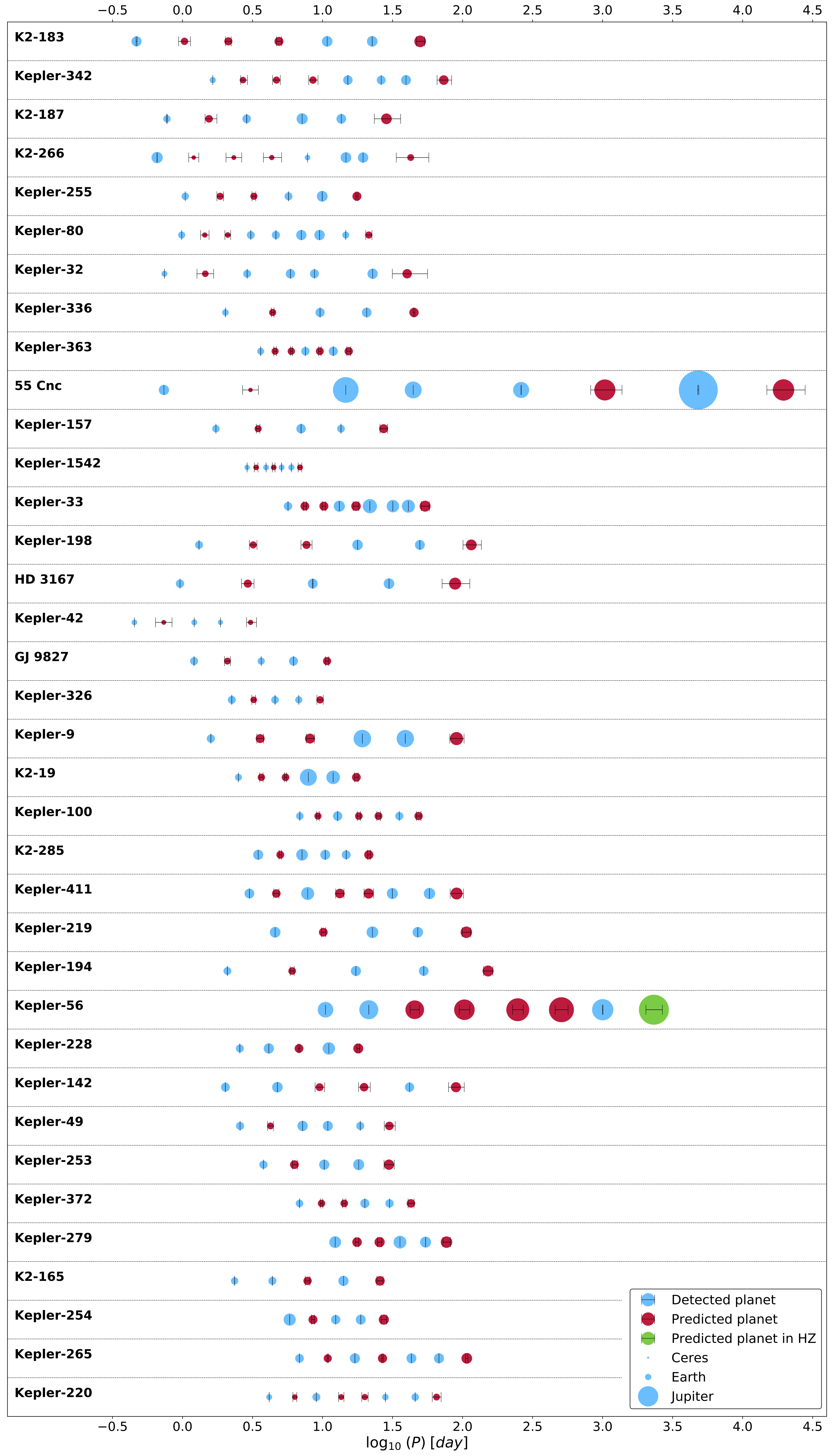}
 \caption{Orbital periods and scaled radii of exoplanets in multiple-planet systems with interpolated and extrapolated planet predictions. The cyan and red circles indicate detected and predicted planets in systems, respectively. The green circles also indicate the predicted planets within the HZ of their parent stars. Due to the discovery methods of KIC 10001893 and PSR B1257+12, the predicted planets' radii are not calculated. Therefore, KIC 10001893 and PSR B1257+12 are excluded.}
 \label{figure8}
\end{figure*}

\begin{figure*}
\ContinuedFloat
\begin{center}
 \includegraphics[width=0.56\paperwidth]{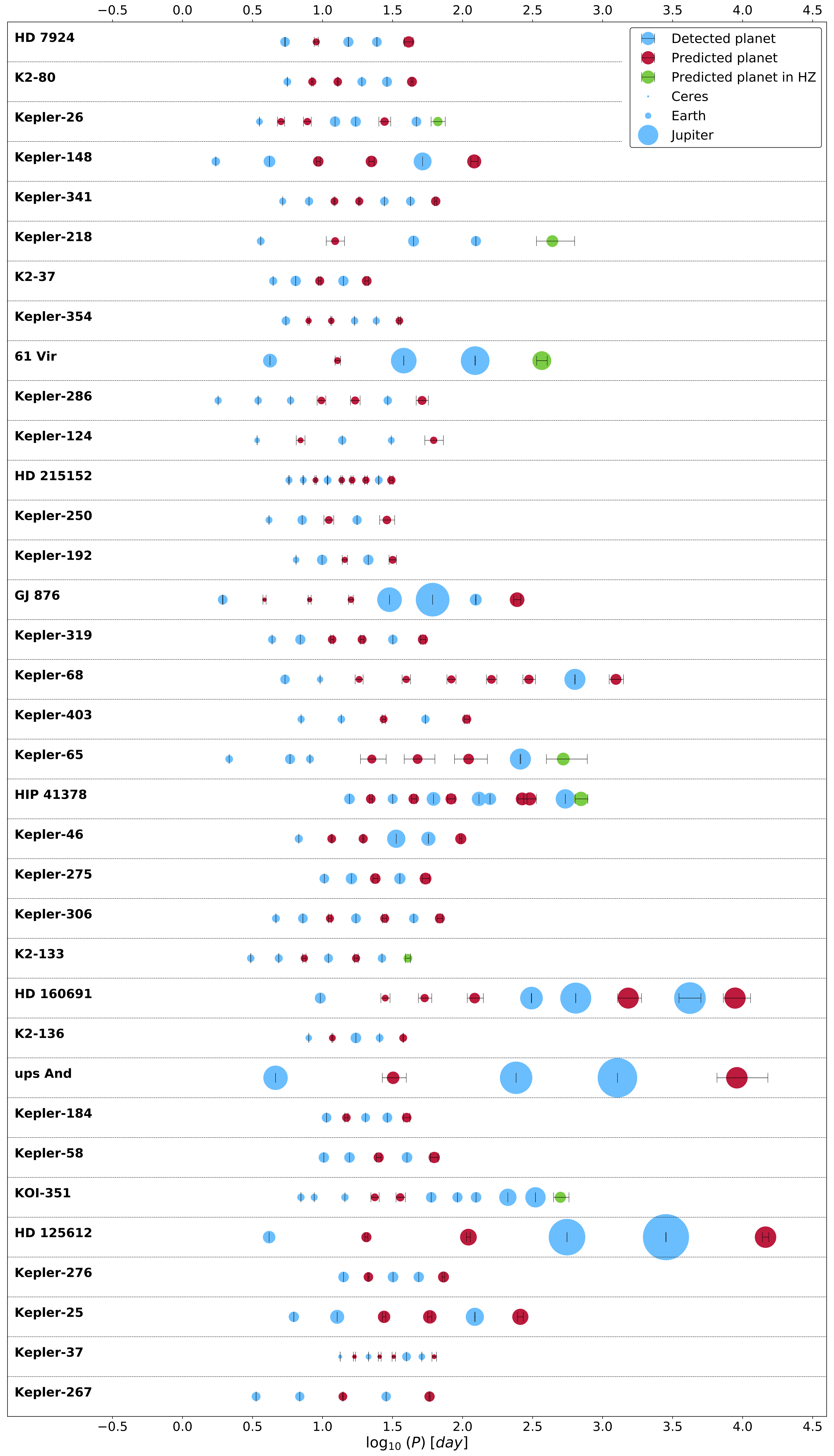}
 \end{center}
 \caption[]{continued}
 \label{}
\end{figure*}

\begin{figure*}
\ContinuedFloat
\begin{center}
 \includegraphics[width=0.56\paperwidth]{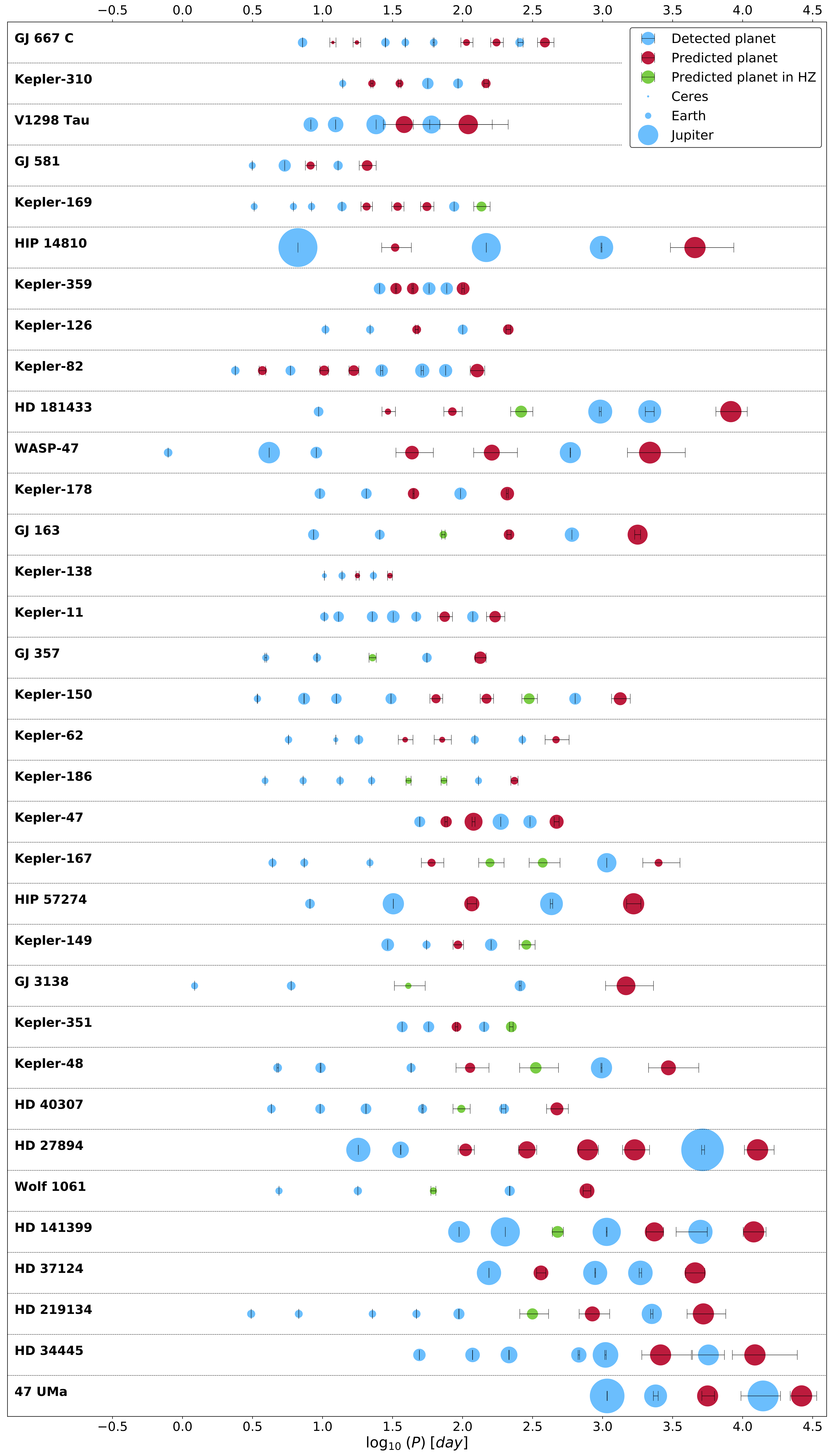}
 \end{center}
 \caption[]{continued}
 \label{}
\end{figure*}

\begin{figure}
 \includegraphics[width=\columnwidth]{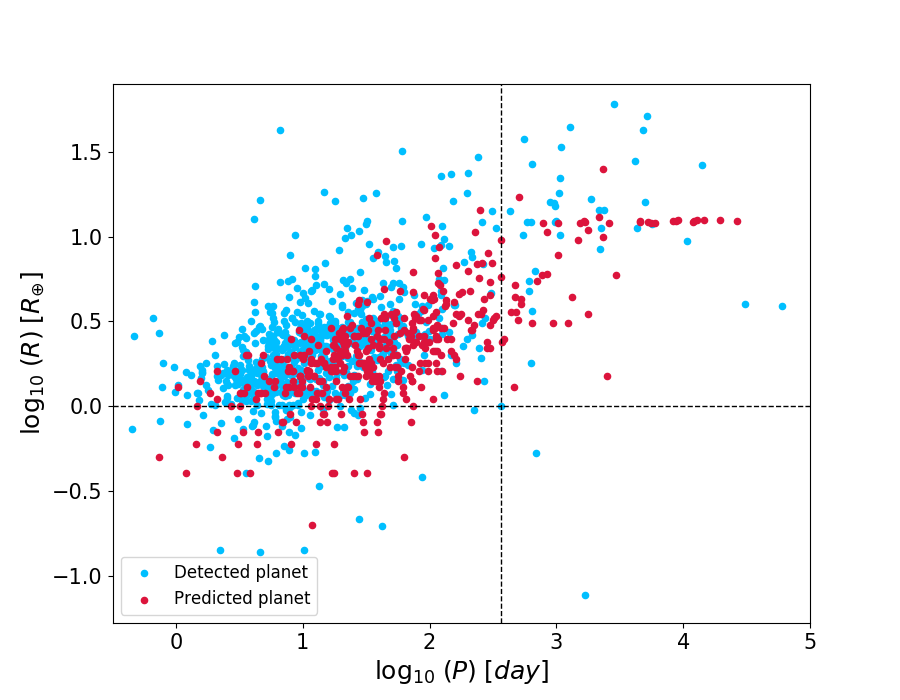}
 \caption{The radius vs. period of detected (blue dots) and predicted (red dots) exoplanets of 229 multi-planetary systems used in this study. The dashed horizontal and vertical lines correspond to the radius and orbital period of the Earth respectively. The vast majority of the detected and predicted planets are in larger radii and tighter orbits than Earth.}
 \label{figure9}
\end{figure}

\begin{table*}
\small
\centering
\caption{Predicted exoplanets within the HZ of host stars in multi-planetary systems. Columns 1, 2, and 3 present the id, host star name, and discovery method (Dis.). Columns 4, 5, and 6 present the orbital period in days, distance from the parent star in AU, and the orbital number (ON). The estimated maximum radius ($R_{Max}$) and maximum mass ($M_{Max}$) in the Earth unit are presented in columns 7 and 8. Column 9 lists the transit probability ($P_{tr}$). The conservative ($HZ_{Cons}$) and optimistic ($HZ_{Opt}$) HZ limits in AU are presented in columns 10 and 11, respectively.}
\label{table2}
\centering
\begin{tabular}{lllllllllll}
\hline \hline
    Panel & Host name & Dis.$^{a}$ & \makecell{Period\\(days)} & \makecell{a\\(AU)} & ON$^b$ & \makecell{$R_{Max}$\\($R_{\oplus}$)} & \makecell{$M_{Max}$\\($M_{\oplus}$)} & \makecell{$P_{tr}$\\(\%)} & \makecell{$HZ_{Cons}$\\(AU)} & \makecell{$HZ_{Opt}$\\(AU)} \\
    \hline
1     & GJ 163 & RV    & $72.9\substack{+2.1\\-2}$ & $0.25\substack{+0.01\\-0.01}$ & 2     & 1.4   & 2.5   & 2.05  & 0.14-0.28 & 0.11-0.30 \\
2     & YZ Cet & RV    & $7.2\substack{+0.4\\-0.3}$ & $0.04\substack{+0.01\\-0.01}$ & 3 E   & 1.2   & 1.9   & 1.96  & 0.05-0.10 & 0.04-0.10 \\
3     & Kepler-445 & Tr    & $13.4\substack{+0.7\\-0.6}$ & $0.06\substack{+0.01\\-0.01}$ & 3 E   & 1.4   & 2.5   & 1.61  & 0.07-0.13 & 0.05-0.14 \\
4     & TOI-270 & Tr    & $20.4\substack{+8.1\\-5.8}$ & $0.11\substack{+0.02\\-0.02}$ & 3 E   & 2.0   & 4.5   & 1.59  & 0.14-0.26 & 0.11-0.28 \\
5     & GJ 357 & RV    & $22.7\substack{+1.4\\-1.3}$ & $0.11\substack{+0.01\\-0.01}$ & 2     & 1.3   & 2.1   & 1.42  & 0.13-0.25 & 0.10-0.27 \\
6     & Kepler-249 & Tr    & $33.1\substack{+0.2\\-0.2}$ & $0.16\substack{+0.01\\-0.01}$ & 3 E   & 1.9   & 4.3   & 1.38  & 0.19-0.37 & 0.15-0.39 \\
7     & Kepler-186 & Tr    & $41.2\substack{+1.7\\-1.7}$ & $0.19\substack{+0.01\\-0.01}$ & 4     & 0.9   & 1.1   & 1.26  & 0.24-0.46 & 0.19-0.49 \\
8     &       &       & $73.6\substack{+3.5\\-3.5}$ & $0.28\substack{+0.01\\-0.01}$ & 5     & 1.0   & 1.4   & 0.85  &       &  \\
9     & K2-133 & Tr    & $40.7\substack{+1.7\\-1.6}$ & $0.18\substack{+0.01\\-0.01}$ & 6 E   & 1.9   & 4.4   & 1.18  & 0.19-0.36 & 0.15-0.38 \\
10    & GJ 3138 & RV    & $41.0\substack{+11.4\\-9.4}$ & $0.20\substack{+0.01\\-0.03}$ & 2     & 1.0   & 1.2   & 1.12  & 0.22-0.42 & 0.17-0.44 \\
11    & K2-72 & Tr    & $39.6\substack{+18\\-13.7}$ & $0.15\substack{+0.04\\-0.04}$ & 4 E   & 1.5   & 2.7   & 1.03  & 0.12-0.23 & 0.09-0.24 \\
12    & Kepler-1388 & Tr    & $74.2\substack{+14.6\\-14.0}$ & $0.30\substack{+0.03\\-0.03}$ & 4 E,C & 2.9   & 9.3   & 0.97  & 0.31-0.59 & 0.25-0.63 \\
13    & Kepler-26 & Tr    & $66.6\substack{+8.1\\-7.4}$ & $0.26\substack{+0.02\\-0.02}$ & 7 E   & 2.2   & 5.7   & 0.89  & 0.24-0.46 & 0.19-0.48 \\
14    & Kepler-52 & Tr    & $77.8\substack{+12.9\\-10.4}$ & $0.29\substack{+0.03\\-0.03}$ & 3 E   & 2.4   & 6.4   & 0.89  & 0.31-0.58 & 0.25-0.61 \\
15    & HD 40307 & RV    & $98.0\substack{+14.3\\-13.4}$ & $0.38\substack{+0.04\\-0.03}$ & 4     & 1.6   & 3.1   & 0.87  & 0.48-0.86 & 0.38-0.91 \\
16    & HD 20781 & RV    & $200.4\substack{+79.7\\-55.4}$ & $0.59\substack{+0.15\\-0.11}$ & 4 E   & 6.3   & 38.0  & 0.85  & 0.69-1.23 & 0.54-1.30 \\
17    & Kepler-331 & Tr    & $63.5\substack{+13.1\\-11.7}$ & $0.27\substack{+0.04\\-0.03}$ & 3 E   & 1.9   & 4.5   & 0.83  & 0.28-0.53 & 0.22-0.55 \\
18    & Kepler-55 & Tr    & $105.6\substack{+58.3\\-37}$ & $0.37\substack{+0.13\\-0.09}$ & 5 E   & 2.8   & 8.6   & 0.77  & 0.38-0.71 & 0.30-0.74 \\
19    & K2-3  & Tr    & $97.4\substack{+71.8\\-42.8}$ & $0.35\substack{+0.15\\-0.11}$ & 3 E   & 1.8   & 4.1   & 0.74  & 0.26-0.50 & 0.21-0.53 \\
20    & K2-155 & Tr    & $98.4\substack{+75.2\\-42.1}$ & $0.36\substack{+0.17\\-0.11}$ & 3 E   & 2.4   & 6.5   & 0.74  & 0.32-0.60 & 0.25-0.63 \\
21    & Kepler-169 & Tr    & $136.5\substack{+19.4\\-17.4}$ & $0.50\substack{+0.05\\-0.04}$ & 8 E   & 2.5   & 7.2   & 0.70  & 0.56-1.02 & 0.45-1.08 \\
22    & Wolf 1061 & RV    & $61.9\substack{+2.5\\-2.5}$ & $0.20\substack{+0.01\\-0.01}$ & 2     & 1.2   & 1.8   & 0.70  & 0.10-0.21 & 0.08-0.22 \\
23    & Kepler-235 & Tr    & $113.3\substack{+13.2\\-11}$ & $0.38\substack{+0.03\\-0.02}$ & 4 E   & 2.5   & 7.1   & 0.66  & 0.30-0.57 & 0.24-0.60 \\
24    & Kepler-167 & Tr    & $157.0\substack{+36.3\\-29.2}$ & $0.52\substack{+0.08\\-0.07}$ & 4     & 2.0   & 4.6   & 0.64  & 0.52-0.94 & 0.41-0.99 \\
25    &       &       & $373.5\substack{+106.6\\-83.3}$ & $0.93\substack{+0.17\\-0.14}$ & 5     & 2.4   & 6.8   & 0.36  &       &  \\
26    & Kepler-166 & Tr    & $163.1\substack{+39.9\\-28}$ & $0.56\substack{+0.09\\-0.07}$ & 3 E   & 3.5   & 13.2  & 0.61  & 0.63-1.12 & 0.50-1.19 \\
27    & Kepler-296 & Tr    & $113.6\substack{+8.2\\-7.2}$ & $0.36\substack{+0.02\\-0.01}$ & 5 E   & 2.1   & 5.0   & 0.61  & 0.21-0.40 & 0.16-0.42 \\
28    & Kepler-351 & Tr    & $223.0\substack{+6.6\\-6.6}$ & $0.69\substack{+0.02\\-0.02}$ & 4 E   & 3.0   & 10.2  & 0.57  & 0.78-1.38 & 0.61-1.45 \\
29    & HD 141399 & RV    & $477.4\substack{+45.5\\-41.5}$ & $1.22\substack{+0.08\\-0.07}$ & 2     & 3.6   & 13.5  & 0.55  & 1.12-1.98 & 0.88-2.09 \\
30    & HD 181433 & RV    & $261.8\substack{+50.5\\-44.9}$ & $0.69\substack{+0.08\\-0.08}$ & 3     & 3.7   & 14.5  & 0.53  & 0.57-1.04 & 0.45-1.09 \\
31    & Kepler-149 & Tr    & $285.5\substack{+41.1\\-33.5}$ & $0.82\substack{+0.08\\-0.06}$ & 4 E   & 2.4   & 6.8   & 0.53  & 0.80-1.43 & 0.63-1.51 \\
32    & Kepler-251 & Tr    & $252.6\substack{+232.8\\-119.5}$ & $0.77\substack{+0.42\\-0.27}$ & 4 E   & 3.5   & 13.1  & 0.53  & 0.79-1.40 & 0.62-1.47 \\
33    & HD 20794 & RV    & $322.7\substack{+155\\-98.2}$ & $0.82\substack{+0.24\\-0.18}$ & 4 E   & 3.3   & 11.9  & 0.52  & 0.91-1.63 & 0.72-1.72 \\
34    & Kepler-56 & Tr    & $2324.8\substack{+325.9\\-305.1}$ & $3.77\substack{+0.34\\-0.34}$ & 7 E   & 25.1  &   -    & 0.52  & 2.97-5.41 & 2.34-5.70 \\
35    & Kepler-150 & Tr    & $298.9\substack{+40.5\\-35.9}$ & $0.87\substack{+0.07\\-0.08}$ & 6     & 3.0   & 10.0  & 0.50  & 0.84-1.49 & 0.66-1.57 \\
36    & Kepler-30 & Tr    & $310.2\substack{+108.4\\-81.9}$ & $0.89\substack{+0.2\\-0.16}$ & 3 E   & 7.0   & 46.6  & 0.49  & 0.83-1.48 & 0.66-1.56 \\
37    & Kepler-65 & Tr    & $524.0\substack{+206.9\\-145.9}$ & $1.37\substack{+0.34\\-0.27}$ & 7 E   & 4.3   & 19.2  & 0.47  & 1.51-2.63 & 1.19-2.78 \\
38    & 61 Vir & RV    & $368.1\substack{+32.8\\-31.3}$ & $0.98\substack{+0.06\\-0.05}$ & 4 E   & 9.6   & 82.2  & 0.45  & 0.86-1.53 & 0.68-1.61 \\
39    & Kepler-48 & Tr    & $333.4\substack{+124.5\\-88.6}$ & $0.90\substack{+0.21\\-0.17}$ & 4     & 3.4   & 12.5  & 0.45  & 0.71-1.27 & 0.56-1.34 \\
40    & Kepler-298 & Tr    & $202.2\substack{+317.5\\-113.7}$ & $0.59\substack{+0.51\\-0.25}$ & 3 E   & 3.2   & 11.0  & 0.45  & 0.35-0.65 & 0.28-0.69 \\
41    & Kepler-218 & Tr    & $437.8\substack{+160.6\\-114.6}$ & $1.14\substack{+0.26\\-0.21}$ & 4 E   & 3.6   & 14.1  & 0.43  & 0.93-1.65 & 0.73-1.74 \\
42    & HD 219134 & RV    & $315.5\substack{+83.7\\-65.9}$ & $0.84\substack{+0.15\\-0.12}$ & 5     & 3.2   & 10.9  & 0.42  & 0.52-0.95 & 0.41-1.00 \\
43    & KOI-351 & Tr    & $499.9\substack{+70.4\\-55.6}$ & $1.31\substack{+0.12\\-0.1}$ & 10 E  & 3.2   & 11.2  & 0.42  & 1.24-2.17 & 0.98-2.29 \\
44    & Kepler-401 & Tr    & $640.4\substack{+299.4\\-188.2}$ & $1.54\substack{+0.46\\-0.32}$ & 3 E   & 3.1   & 10.6  & 0.40  & 1.39-2.43 & 1.10-2.56 \\
45    & HIP 41378 & Tr    & $701.2\substack{+78.3\\-67.8}$ & $1.63\substack{+0.12\\-0.11}$ & 11 E  & 5.5   & 30.2  & 0.38  & 1.44-2.51 & 1.14-2.65 \\
46    & Kepler-603 & Tr    & $527.1\substack{+1408.5\\-356.9}$ & $1.28\substack{+1.77\\-0.68}$ & 3 E   & 4.0   & 16.8  & 0.36  & 0.97-1.71 & 0.76-1.80 \\
47    & HD 136352 & RV    & $302.2\substack{+462.8\\-188.2}$ & $0.82\substack{+0.71\\-0.39}$ & 3 E   & 5.4   & 28.6  & -     & 0.95-1.69 & 0.75-1.78 \\
    \hline \hline
    \end{tabular}
\medskip
  \tabnote{$^{a}$Discovery method of the system: 'Tr' and 'RV' represent transit and radial velocity, respectively.}
  \tabnote{$^b$Orbital numbers (ON) followed by 'E' indicate the extrapolated planets, and followed by 'C' indicate that the corresponding orbital periods have been flagged as "Planetary Candidate" in the NASA Exoplanet Archive.}
\end{table*}

\section{RESULTS}\label{results}
\subsection{The prediction of additional exoplanets in exoplanetary systems}
To investigate the probability of the existence of additional planets in exoplanetary systems, we apply the TB relation and the MCMC method to analyse the data of a sample of 229 systems that contain at least three confirmed exoplanets. We find that 122 systems adhere to the TB relation better than the Solar System without any need for interpolation. For those 107 systems that adhere to the TB relation worse than the Solar System, we insert up to 10 new additional planets. For example, Fig. \ref{figure4} shows the best linear regression (TB relation) for up to 10 additional planet inserts in the GJ 667 C system. For GJ 667 C, the highest $\gamma$ value is in the fourth step, where the number of inserted planets ($n_{ins}$) is four. Black dots represent the detected planets and red dots the predicted planets. The black line shows the best (mean) scaling relation, and the two dashed lines show the $\pm1\sigma$ uncertainties around this relation. The blue lines are a set of 100 different realisations, drawn from the multivariate Gaussian distribution of the parameters, where for the highest $\gamma$ value: m=0.1924, b=0.856, and ln<$\sigma$>=-3.73. The scatter covariance matrix is also estimated from the MCMC chain. Figure \ref{figure5} illustrates the one- and two-dimensional marginalised posterior distributions of the scaling relation parameters for the fourth step of the linear regression for the GJ 667  C system. 

Table \ref{table1} represents the data corresponding to Fig. \ref{figure4} and lists the number of inserted planets ($n_{ins}$) (column 1), the values of parameters $\chi^2/dof$ (column 2), $\gamma$ for each step of interpolation (column 3), orbital periods (column 4), and orbital numbers (ON) of inserted planets (column 5).

After interpolating all 107 systems, we find that these systems adhere to the TB relation better than the Solar System or to approximately the same extent. Of these 107 interpolated systems, 50 systems need one, 33 systems need two, and the remaining 24 systems need more than two additional planets to be inserted. We also predict an extrapolated planet beyond the outermost detected planet for all systems in our sample. We predict the existence of 426 possible additional exoplanets in these systems, of which 197 are predicted by interpolation. It should be noted that six of the predicted planets in Kepler-1388, Kepler-1542, Kepler-164, Kepler-374, Kepler-402, and Kepler-403 have been flagged as "Planetary Candidates" in the NASA Exoplanet Archive.

To verify whether some of the predicted exoplanets are in the HZ of their parent stars, we use conservative and optimistic definitions of the HZ \citep{2012PASP..124..323K}. Among the predicted exoplanets, 47 exoplanets lie within the HZ of their host stars. 27 exoplanets out of 47 are located within the conservative HZ. Furthermore, 14 exoplanets in the HZ have been predicted by interpolation, and the remaining 33 exoplanets have been predicted by extrapolation. Table \ref{table2} presents the 47 predicted exoplanets that lie within the HZ of the host star. Columns 1, 2, and 3 present the id, the host star name, and discovery method (Dis.), respectively. Columns 4, 5, and 6 present the orbital period in days, the distance from the parent star (a) in AU, and the orbital number (ON). The estimated maximum radius and maximum mass in the Earth unit are presented in columns 7 and 8. Column 9 lists the transit probability ($P{tr}$). The conservative and optimistic HZ limits for AU are presented in columns 10 and 11, respectively. Kepler-167 is a four-planet system where we predict three interpolated additional exoplanets, in which two planets have an orbital period of 157.0 and 373.5 days, located within the conservative HZ. The host star Kepler-186 is also a five-planet system including two predicted exoplanets with orbital periods of 41.2 and 73.6 days; these two additional planets were found by interpolation and are located within the optimistic and conservative HZ, respectively.

\citet{2011ApJ...736...19B} classified exoplanets into the following class sizes: Earth-size ($R_{p}<1.25R_{\oplus}$), super-Earth-size ($1.25R_{\oplus}\leq R_{p}<2R_{\oplus}$), Neptune-size ($2R_{\oplus}\leq R_{p}<6R_{\oplus}$) and Jupiter-size ($6R_{\oplus}\leq R_{p}<15R_{\oplus}$). Following this classification, five of our predicted exoplanets within HZ have maximum radii within the Earth-size range, 11 super-Earth-size, 27 Neptune-size, three Jupiter-size, and one with a maximum radius larger than twice that of Jupiter's. Using the proposed categories based on planet mass by \citet{2013PASP..125..933S}, our five predicted planets have maximum masses within the range of Earth ($0.1M_{\oplus}-2M_{\oplus}$). In addition, 22 and 19 predicted planets have maximum masses within the range of super-Earth ($2M_{\oplus}-10M_{\oplus}$) and Neptune ($10M_{\oplus}-100_{\oplus}$), respectively. As a result, among our 47 predicted exoplanets within HZ, there are only five exoplanets whose estimated maximum mass and radius are within the mass and radius range of Earth: the fourth and fifth planets of Kepler-186, the second planets of GJ 3138 and Wolf 1061, and the extrapolated planet of YZ Cet.

We use the dynamical spacing criterion ($\Delta$) to investigate our predicted objects' stability at HZ. We calculate the $\Delta$ values for all adjacent planet pairs in these 45 systems, which host 47 predicted exoplanets within their HZ. We find that when inserting our predicted planets into systems, the average percentage of pairs with $\Delta\leq10$ increases from $\sim25\%$ to $\sim38\%$. Nevertheless, this alone cannot be a reason for instability. In our Solar System, there are pairs of objects whose dynamical spacing values are small ($\Delta\leq10$); however, they are stable in their positions. Neptune and Pluto are one of those pairs; the dynamical spacing value for them is $\Delta\sim7.4$. We know that Neptune and Pluto are stable because they are in a 3:2 orbital resonance with each other. The same could be true for exoplanetary systems as found by \citet{2012ApJ...756L..11L}, \citet{2013AJ....145....1B}, and \citet{2014ApJ...790..146F}. We calculate the ratio of the orbital periods of planet pairs in these 45 systems and find that the number of orbital resonances increases when our predicted planets are considered. Figure \ref{figure6} shows the number of resonance pairs with and without considering our predicted planets as shown with the solid blue line and dotted black line, respectively. The values for the Solar System are also shown with reference to the orange dashed line. As shown in this figure, the $\Delta\leq10$ pairs are dominated by the pairs in resonance; hence, as a sample, the planets (both detected and predicted) of these 45 systems are more likely to be stable in their positions.

Table \ref{table3} lists our predicted exoplanets by extrapolation and the best-fit TB relations. Columns 1 and 2 present the host star name and the system discovery method (Dis.), respectively. Column 3 reports a flag that defines whether the system has already been analysed by \citetalias{2015MNRAS.448.3608B} (or \citetalias{2013MNRAS.435.1126B}) (Y) or not (N). Columns 4 to 7 present $\chi^2/dof$, slope (m), intercept (b), and predicted orbital period, respectively. Column 8 reports whether the predicted period values in this paper and \citetalias{2015MNRAS.448.3608B} (or \citetalias{2013MNRAS.435.1126B}) are consistent within error (Y) or not (N). Columns 9 to 12, respectively, list the orbital number (ON), estimated maximum radius ($R_{Max}$), and maximum mass ($M_{Max}$) in the Earth radius and mass unit, and the transit probability ($P_{tr}$). The columns have been sorted based on the transit probability in descending order.

Table \ref{table4} lists the systems with interpolated and extrapolated planet predictions. In this table, we present the $\chi^2/dof$ before and after interpolation in the fourth and fifth columns. $\gamma$ and $\Delta\gamma$ (where $\Delta\gamma$=($\gamma_{1}$-$\gamma_{2}$)/$\gamma_{2}$; $\gamma_{1}$ and $\gamma_{2}$ are the highest and second-highest $\gamma$ values for system, respectively) are listed in columns 6 and 7, respectively. The definitions of other columns in this table are the same as in Tab. \ref{table3}. Table \ref{table4} is also sorted by the transit probability in descending order.

In Fig. \ref{figure7}, we show the new systems containing the detected planets (cyan circles) and the predicted (red circles) planets by extrapolation, while the predicted planets within the HZ are also shown as green circles. The sizes of the symbols are scaled based on the planet's radius. Similarly, Fig. \ref{figure8} presents the detected exoplanets (cyan circles) and the predicted exoplanets (red circles) by interpolation and extrapolation.

Figure \ref{figure9} illustrates the radius versus the orbital period of the detected (blue points) and predicted (red points) exoplanets. We find similar trends for both detected and predicted exoplanets. As seen, the vast majority of exoplanets have larger radii and shorter orbital periods than Earth (the radius and orbital period of the Earth have been shown with vertical and horizontal dotted black lines) due to observational limitations.

We note that the estimated masses (and radii) of GJ 676 A, Kepler-56, and WASP-47 are excluded because they are higher than the maximum possible limit of a typical planet. Furthermore, due to the lack of stellar parameters, the transit probabilities of HD 31527, HD 136352, and Kepler-402 are not calculated.

\subsection{The reliability of planet predictions using the TB relation}
To examine the reliability of the predictions made by the TB relation, we look for some planetary systems in the literature that have had new exoplanets detected recently, notably those systems that have been predicted to have additional planets by \citetalias{2015MNRAS.448.3608B}. We find the following seven systems with new planet detection: Kepler-1388, Kepler-150, Kepler-1542, Kepler-20, Kepler-80, Kepler-82, and KOI-351.

The detected planets in Kepler-150 and Kepler-82 were not predicted by \citetalias{2015MNRAS.448.3608B} \citep{2017AJ....153..180S,2019A&A...628A.108F}, while the rest of the five planets have been detected with orbital periods that agree with their predicted periods. For Kepler-1388 and Kepler-1542, two planetary candidates with orbital periods of 75.73 and 7.23 days have been detected, the detected periods of which are consistent with their predicted orbital periods, $73.0\pm8.0$ and $6.9\pm0.3$ days, by \citetalias{2015MNRAS.448.3608B} and also with our predictions $74.2\pm14.3$ and $6.9\pm0.2$ days. \citetalias{2015MNRAS.448.3608B} could also predict an additional planet in Kepler-20 with an orbital period of $39.1\pm5.4$ days. This prediction was also confirmed by \citet{2016AJ....152..160B}. \citet{2018AJ....155...94S} used deep learning algorithms to classify potential planet signals, where they managed to validate two more new planets in Kepler-80 and KOI-351, which were previously predicted by \citetalias{2015MNRAS.448.3608B}. Table \ref{table5} summarises these results and presents the predicted and detected periods, which are reasonably consistent, especially for Kepler-1542 and KOI-351 where \citetalias{2015MNRAS.448.3608B} highlighted them as predictions with a high geometric probability to transit. This demonstrates the potential and capability of such predictions to help search for new planets, using more precise observational data and new detection techniques such as machine learning. However, having seven systems is not statistically significant enough to establish either the reliability or unreliability of the TB relation, and we require more follow-up observations. We must wait for ongoing exoplanet surveys such as TESS to be completed.

\subsection{Comparison between predictions by \citetalias{2015MNRAS.448.3608B} and this study}
This study uses the TB relation to predict the probability of finding additional planets in multiplanet systems with at least three member planets, similar to the method used by \citetalias{2015MNRAS.448.3608B}. In principle, we update the \citetalias{2015MNRAS.448.3608B} study and utilise the Markov Chain Monte Carlo (MCMC) Simulation. The MCMC method is used to analyse systems and quantify the uncertainties of the best-fit parameters \citep{emcee3}. The main feature of our method is inserting thousands of hypothetical planets with random orbital periods into each of the systems and using the highest $\gamma$ value to achieve the most accurate predictions. 

To compare our method and that of \citetalias{2015MNRAS.448.3608B}, we remove Ceres and the planet Uranus (two objects predicted by the TB law to exist) from the Solar System and insert thousands of random planets into the Solar System, covering all possible locations and combinations between the two adjacent planets, to achieve the minimum value of $\chi^2/dof$. This leads us to recover the actual Solar System with highest $\gamma$ value, where the predicted orbital periods' calculated errors are 20\% for Ceres and 23\% for Uranus. On the other hand, the \citetalias{2015MNRAS.448.3608B} method results in an error value of 23\% for both Ceres and Uranus. We remove the planets Earth and Mars, which are within the HZ range of the Sun, and apply the TB relation to the system to predict their orbital periods. This recovers the combination of the actual Solar System where the calculated errors of orbital periods are 18\% for Earth and 41\% for Mars. We perform this process once again using the \citetalias{2015MNRAS.448.3608B} method, which gives us the error values of 18.5\% for Earth and 42\% for Mars. For a better comparison, we repeat this process for other planets in the Solar System and exoplanetary systems, and with different combinations of planet removal.

\begin{figure}
 \includegraphics[width=\columnwidth]{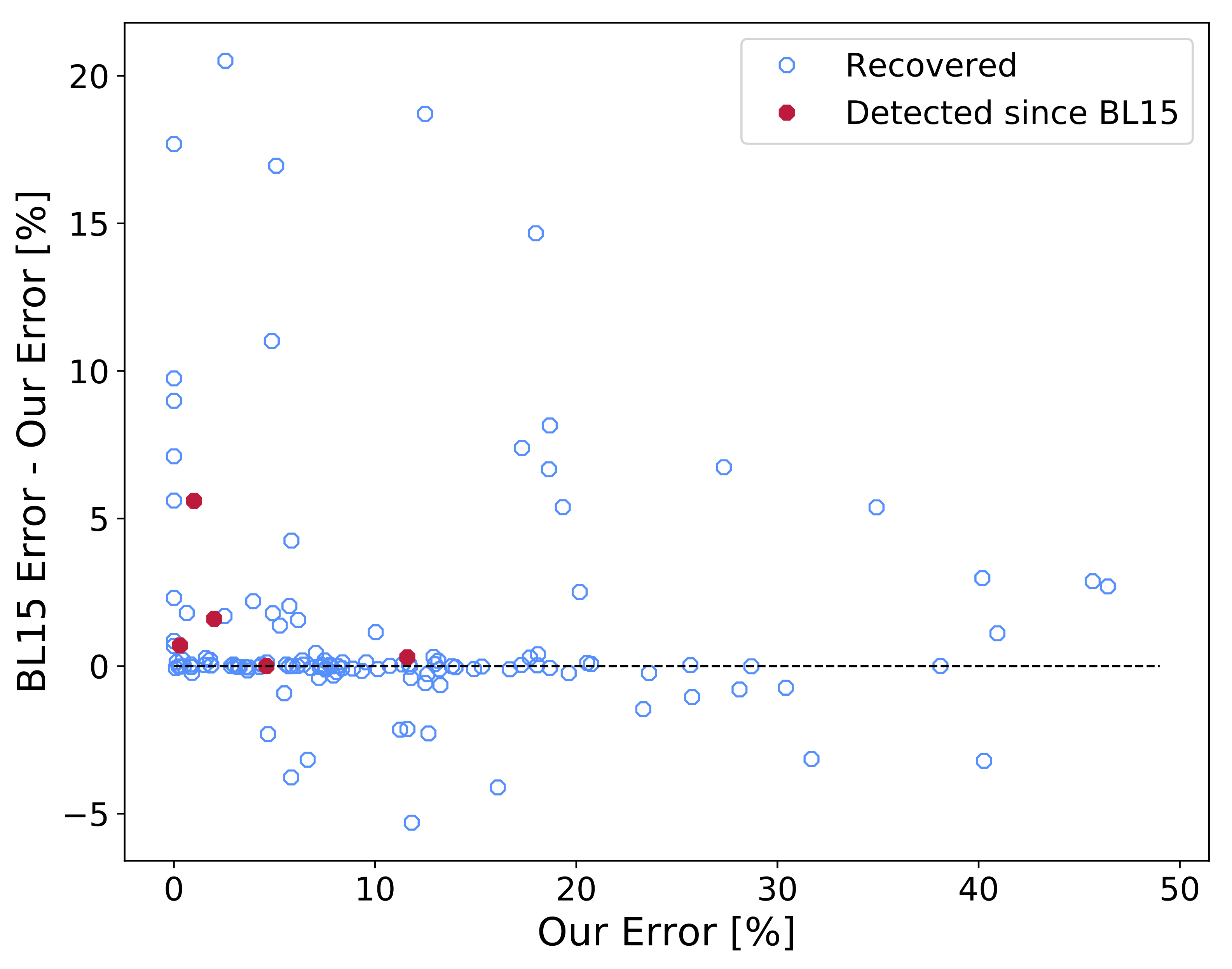}
 \caption{Comparison of uncertainty on the predicted orbital periods calculated by \citetalias{2015MNRAS.448.3608B} and this study. We remove planets from systems in various combinations and apply the TB relation to recover the orbital periods of removed planets. Each blue data point belongs to a specific combination of removed planets from systems, and five red data points represent those detected planets after the predictions (see Tab.  \ref{table5}) made by \citetalias{2015MNRAS.448.3608B}.}
 \label{figure10}
\end{figure}

We also reapply our method to Kepler-1388, Kepler-1542, Kepler-20, Kepler-80, and KOI-351, regardless of the successfully prediction of planets by \citetalias{2015MNRAS.448.3608B}, to compare the validity of the methods used in predicting the additional planets. As shown in Tab. \ref{table5}, our predictions have fewer errors, and sometimes the predicted periods have smaller error bars. These fewer error bars can be interpreted as advantages of using the MCMC method in predicting exoplanets based on the TB relation. Using the paired samples t-test, the p-value is estimated to equal 0.025, which is less than 0.05 and statistically significant. Therefore, we conclude that, on average, there is evidence that applying the MCMC method does lead to more precise predictions than \citetalias{2015MNRAS.448.3608B}'s method. The difference between the calculated errors with our method and \citetalias{2015MNRAS.448.3608B}'s is illustrated in Fig. \ref{figure10}, where each blue data point belongs to a planet recovery. Five red data points represent those detected planets after the predictions made by \citetalias{2015MNRAS.448.3608B}.

\newpage
\section{SUMMARY AND CONCLUSIONS}\label{summary}
Using the Markov Chain Monte Carlo method, we apply the TB relation to all available exoplanetary systems with three or more confirmed exoplanets, a total of 229 systems, to examine their adherence to the TB relation in comparison to the Solar System and to predict the existence of possible additional planets. For those systems that adhere to the TB relation better than the Solar System, we extrapolate one additional planet, and for each of the remaining systems that adhere worse, we interpolate up to 10 specific planets between the detected planets and identify new possible additional planets in the systems.

We present a list of 229 analysed exoplanetary systems (of which 123 of them have not been previously analysed by either \citetalias{2013MNRAS.435.1126B} or \citetalias{2015MNRAS.448.3608B}) containing their unique TB relations and a total of 426 additional predicted exoplanets, of which 47 are located within the HZ of their parent stars. We also estimate that five of the predicted planets in HZ have maximum mass and radius limits within the Earth's mass and radius range. 

As an important result, the planets of $\sim53\%$ of our sample system adhere to a logarithmic spacing relation better than the planets of the Solar System. Therefore, there is a need to work with more comprehensive data of multiplanetary systems to reveal the probable dynamical or gravitational aspects of the TB relation.

We find new planet detection for seven exoplanetary systems after the predictions made by \citetalias{2015MNRAS.448.3608B} and compare the detected and predicted orbital periods. We find that both the detected and predicted orbital periods agree very well within errors. Our predictions also agree roughly better than those made by \citetalias{2015MNRAS.448.3608B}, indicating that using a precise modeling method and measurements could improve our predictions and uncertainties as well. However, to claim the (un)reliability of the TB relation in predicting the presence of additional planets in exoplanetary systems, we require much more data and further follow-up observations of the exoplanetary systems. Thus, we must wait for upcoming new exoplanet surveys or ongoing surveys such as TESS.

\begin{acknowledgements}
This research has used theNASA Exoplanet Archive, Extrasolar Planets Encyclopedia, and Habitable Zone Gallery. We wish to thank Kenneth P. K. Quek for the English language editing. We also thank Ashkan M. Jasour and  Thomas Hackman for their useful comments on the manuscript. 
The author acknowledges the usage of the following python packages, in alphabetical order: \texttt{astropy} \citep{astro2013,astro2018}, \texttt{chainConsumer} \citep{2019ascl.soft10017H}, 
\texttt{emcee} \citep{emcee3}, \texttt{matplotlib} \citep{mathlib}, \texttt{numpy} \citep{numpy2}, and \texttt{scipy} \citep{scipy}.
\end{acknowledgements}

\begin{table*}
\small
\centering
\caption{Systems with only extrapolated planet predictions. Columns 1 and 2 present the host star name and discovery method (Dis.). Column 3 reports a flag that defines whether the system has already been analysed by \citetalias{2015MNRAS.448.3608B} (or \citetalias{2013MNRAS.435.1126B}) (Y) or not (N). Columns 4 to 7 present $\chi^2/dof$, slope (m), intercept (b), and predicted orbital period, respectively. Column 8 reports whether the predicted period values in this paper and \citetalias{2015MNRAS.448.3608B} (or \citetalias{2013MNRAS.435.1126B}) are consistent within error (Y) or not (N). Columns 9 to 12, respectively, list the orbital number (ON), estimated maximum radius ($R_{Max}$), and maximum mass ($M_{Max}$) in the Earth radius and mass unit, and the transit probability ($P_{tr}$).}
\label{table3}
    \begin{tabular}{llllllllllll}
    \hline \hline
    Host name & Dis.$^{a}$ & $F_{1}$ & $\frac{\chi^2}{dof}$ & m & b     & \makecell{Period\\(days)} & $F_{2}$ & ON$^{b}$    & \makecell{$R_{Max}$\\($R_{\oplus}$)} & \makecell{$M_{Max}$\\($M_{\oplus}$)} & \makecell{$P_{tr}$\\(\%)} \\
    \hline
Solar System & -     & Y     & 1.000 & $0.358\substack{+0.011\\-0.012}$ & $1.885\substack{+0.053\\-0.056}$ & -     & -     & -     & -     & -     & - \\
Kepler-207 & Tr    & Y     & 0.001 & $0.281\substack{+0.001\\-0.001}$ & $0.207\substack{+0.001\\-0.001}$ & $11.2\substack{+0.1\\-0.1}$ & Y     & 3     & 2.1   & 5.2   & 7.31 \\
Kepler-217 & Tr    & Y     & 0.676 & $0.174\substack{+0.036\\-0.027}$ & $0.579\substack{+0.039\\-0.045}$ & $12.6\substack{+5.1\\-3.2}$ & Y     & 3     & 1.8   & 3.9   & 6.90 \\
Kepler-374 & Tr    & Y     & 0.320 & $0.213\substack{+0.033\\-0.021}$ & $0.284\substack{+0.030\\-0.040}$ & $8.4\substack{+2.9\\-1.8}$ & N     & 3 C   & 1.4   & 2.5   & 5.23 \\
Kepler-60 & Tr    & Y     & 0.327 & $0.111\substack{+0.017\\-0.014}$ & $0.849\substack{+0.020\\-0.018}$ & $15.2\substack{+2.7\\-2}$ & Y     & 3     & 2.1   & 5.2   & 4.83 \\
Kepler-23 & Tr    & Y     & 0.135 & $0.166\substack{+0.012\\-0.012}$ & $0.857\substack{+0.017\\-0.013}$ & $22.7\substack{+2.9\\-2.5}$ & Y     & 3     & 2.4   & 6.6   & 4.46 \\
Kepler-223 & Tr    & Y     & 0.318 & $0.146\substack{+0.008\\-0.009}$ & $0.864\substack{+0.017\\-0.016}$ & $28\substack{+3.5\\-3.2}$ & Y     & 4     & 4.2   & 18.3  & 4.23 \\
Kepler-431 & Tr    & Y     & 0.306 & $0.121\substack{+0.013\\-0.013}$ & $0.828\substack{+0.019\\-0.018}$ & $15.5\substack{+2.2\\-1.9}$ & Y     & 3     & 0.8   & 0.9   & 4.18 \\
HR 858 & Tr    & N     & 0.234 & $0.250\substack{+0.026\\-0.020}$ & $0.544\substack{+0.025\\-0.035}$ & $19.7\substack{+5.4\\-3.9}$ & N     & 3     & 2.5   & 7.1   & 4.02 \\
Kepler-256 & Tr    & Y     & 0.236 & $0.269\substack{+0.016\\-0.016}$ & $0.233\substack{+0.032\\-0.029}$ & $20.3\substack{+5.1\\-4}$ & Y     & 4     & 2.8   & 8.7   & 3.99 \\
Kepler-107 & Tr    & Y     & 0.410 & $0.220\substack{+0.016\\-0.016}$ & $0.488\substack{+0.026\\-0.031}$ & $23.3\substack{+5.3\\-4.6}$ & Y     & 4     & 1.1   & 1.6   & 3.92 \\
K2-219 & Tr    & N     & 0.010 & $0.228\substack{+0.005\\-0.003}$ & $0.592\substack{+0.004\\-0.006}$ & $18.9\substack{+0.9\\-0.7}$ & N     & 3     & 1.9   & 4.3   & 3.91 \\
Kepler-226 & Tr    & Y     & 0.474 & $0.157\substack{+0.022\\-0.023}$ & $0.586\substack{+0.030\\-0.030}$ & $11.5\substack{+2.9\\-2.3}$ & Y     & 3     & 1.3   & 2.2   & 3.68 \\
Kepler-271 & Tr    & Y     & 0.001 & $0.149\substack{+0.001\\-0.001}$ & $0.720\substack{+0.001\\-0.001}$ & $14.7\substack{+0.1\\-0.1}$ & N     & 3     & 0.9   & 1.1   & 3.64 \\
Kepler-444 & Tr    & Y     & 0.227 & $0.109\substack{+0.005\\-0.004}$ & $0.558\substack{+0.011\\-0.011}$ & $12.7\substack{+1.1\\-0.9}$ & Y     & 5     & 0.6   & 0.5   & 3.56 \\
Kepler-208 & Tr    & Y     & 0.599 & $0.193\substack{+0.018\\-0.018}$ & $0.650\substack{+0.032\\-0.034}$ & $26.4\substack{+7.3\\-5.7}$ & Y     & 4     & 1.5   & 2.8   & 3.54 \\
Kepler-203 & Tr    & Y     & 0.591 & $0.280\substack{+0.048\\-0.039}$ & $0.483\substack{+0.048\\-0.064}$ & $21\substack{+11.7\\-7.1}$ & Y     & 3     & 1.7   & 3.5   & 3.40 \\
Kepler-758 & Tr    & Y     & 0.215 & $0.208\substack{+0.013\\-0.012}$ & $0.684\substack{+0.021\\-0.027}$ & $32.8\substack{+6\\-5.2}$ & Y     & 4     & 2.2   & 5.6   & 3.11 \\
Kepler-339 & Tr    & Y     & 0.193 & $0.163\substack{+0.015\\-0.014}$ & $0.691\substack{+0.018\\-0.020}$ & $15.2\substack{+2.4\\-2}$ & Y     & 3     & 1.3   & 2.2   & 3.07 \\
Kepler-272 & Tr    & Y     & 0.177 & $0.285\substack{+0.030\\-0.021}$ & $0.480\substack{+0.026\\-0.031}$ & $21.6\substack{+6.5\\-4.2}$ & Y     & 3     & 2.4   & 6.6   & 3.06 \\
Kepler-350 & Tr    & Y     & 0.207 & $0.184\substack{+0.018\\-0.017}$ & $1.054\substack{+0.023\\-0.027}$ & $40.4\substack{+7.9\\-6.7}$ & Y     & 3     & 2.5   & 7.1   & 2.93 \\
K2-148 & Tr    & N     & 0.411 & $0.173\substack{+0.023\\-0.024}$ & $0.652\substack{+0.034\\-0.029}$ & $14.8\substack{+4\\-3.1}$ & N     & 3     & 1.8   & 3.9   & 2.90 \\
Kepler-1254 & Tr    & Y     & 0.171 & $0.224\substack{+0.022\\-0.017}$ & $0.548\substack{+0.024\\-0.027}$ & $16.6\substack{+3.8\\-2.7}$ & Y     & 3     & 1.7   & 3.5   & 2.88 \\
K2-138 & Tr    & N     & 0.007 & $0.183\substack{+0.001\\-0.001}$ & $0.369\substack{+0.003\\-0.003}$ & $19.3\substack{+0.4\\-0.4}$ & N     & 5     & 2.7   & 8.2   & 2.83 \\
Kepler-24 & Tr    & Y     & 0.737 & $0.215\substack{+0.021\\-0.020}$ & $0.655\substack{+0.038\\-0.044}$ & $32.7\substack{+10.6\\-8.1}$ & Y     & 4     & 2.8   & 8.7   & 2.83 \\
Kepler-18 & Tr    & Y     & 0.129 & $0.315\substack{+0.026\\-0.024}$ & $0.552\substack{+0.034\\-0.033}$ & $31.5\substack{+9.3\\-6.8}$ & Y     & 3     & 2.7   & 8.2   & 2.69 \\
Kepler-114 & Tr    & Y     & 0.097 & $0.178\substack{+0.012\\-0.011}$ & $0.718\substack{+0.015\\-0.018}$ & $17.9\substack{+2.3\\-1.9}$ & Y     & 3     & 1.7   & 3.5   & 2.57 \\
Kepler-305 & Tr    & Y     & 0.610 & $0.233\substack{+0.023\\-0.020}$ & $0.499\substack{+0.037\\-0.045}$ & $27.0\substack{+9.2\\-6.8}$ & Y     & 5     & 3.0   & 9.9   & 2.50 \\
Kepler-304 & Tr    & Y     & 0.568 & $0.264\substack{+0.024\\-0.023}$ & $0.205\substack{+0.043\\-0.047}$ & $18.3\substack{+7\\-5}$ & Y     & 4     & 2.2   & 5.6   & 2.44 \\
Kepler-206 & Tr    & Y     & 0.052 & $0.240\substack{+0.016\\-0.013}$ & $0.887\substack{+0.017\\-0.019}$ & $40.5\substack{+6.6\\-5}$ & Y     & 3     & 1.4   & 2.5   & 2.38 \\
Kepler-197 & Tr    & Y     & 0.390 & $0.214\substack{+0.016\\-0.017}$ & $0.769\substack{+0.031\\-0.031}$ & $42.1\substack{+10.2\\-8.5}$ & Y     & 4     & 1.0   & 1.3   & 2.34 \\
Kepler-398 & Tr    & Y     & 0.001 & $0.223\substack{+0.001\\-0.001}$ & $0.611\substack{+0.001\\-0.001}$ & $19.1\substack{+0.1\\-0.1}$ & Y     & 3     & 1.1   & 1.6   & 2.34 \\
Kepler-338 & Tr    & Y     & 0.580 & $0.228\substack{+0.020\\-0.019}$ & $0.942\substack{+0.036\\-0.034}$ & $71.6\substack{+21.6\\-16.3}$ & Y     & 4     & 2.6   & 7.6   & 2.29 \\
Kepler-450 & Tr    & Y     & 0.126 & $0.290\substack{+0.023\\-0.020}$ & $0.883\substack{+0.028\\-0.028}$ & $56.5\substack{+14.1\\-10.3}$ & Y     & 3     & 1.4   & 2.5   & 2.27 \\
Kepler-446 & Tr    & Y     & 0.257 & $0.259\substack{+0.029\\-0.026}$ & $0.204\substack{+0.037\\-0.036}$ & $9.6\substack{+3.2\\-2.2}$ & Y     & 3     & 1.5   & 2.8   & 2.21 \\
Kepler-301 & Tr    & Y     & 0.185 & $0.371\substack{+0.037\\-0.029}$ & $0.384\substack{+0.037\\-0.055}$ & $31.5\substack{+12.7\\-8.8}$ & Y     & 3     & 2.1   & 5.2   & 2.18 \\
K2-239 & Tr    & N     & 0.818 & $0.143\substack{+0.026\\-0.026}$ & $0.727\substack{+0.036\\-0.041}$ & $14.3\substack{+4.3\\-3.5}$ & N     & 3     & 1.2   & 1.9   & 2.07 \\
L 98-59 & Tr    & N     & 0.624 & $0.258\substack{+0.042\\-0.042}$ & $0.341\substack{+0.055\\-0.045}$ & $13\substack{+6.7\\-4.2}$ & N     & 3     & 1.2   & 1.9   & 2.04 \\
Kepler-191 & Tr    & Y     & 0.074 & $0.238\substack{+0.013\\-0.013}$ & $0.768\substack{+0.015\\-0.017}$ & $30.3\substack{+4\\-3.6}$ & Y     & 3     & 1.8   & 3.9   & 2.02 \\
YZ Cet & RV    & N     & 0.012 & $0.187\substack{+0.005\\-0.004}$ & $0.296\substack{+0.006\\-0.006}$ & $7.2\substack{+0.4\\-0.3}$ & N     & 3 H   & 1.2   & 1.9   & 1.96 \\
Kepler-85 & Tr    & Y     & 0.106 & $0.160\substack{+0.005\\-0.005}$ & $0.927\substack{+0.010\\-0.011}$ & $37\substack{+2.6\\-2.6}$ & Y     & 4     & 1.4   & 2.5   & 1.95 \\
Kepler-292 & Tr    & Y     & 0.396 & $0.232\substack{+0.012\\-0.012}$ & $0.381\substack{+0.027\\-0.030}$ & $34.7\substack{+7.8\\-6.4}$ & Y     & 5     & 2.5   & 7.1   & 1.91 \\
Kepler-221 & Tr    & Y     & 0.135 & $0.269\substack{+0.012\\-0.013}$ & $0.463\substack{+0.023\\-0.021}$ & $34.7\substack{+6.3\\-5.4}$ & Y     & 4     & 3.1   & 10.5  & 1.89 \\
K2-198 & Tr    & N     & 0.007 & $0.352\substack{+0.005\\-0.007}$ & $0.524\substack{+0.008\\-0.007}$ & $38.2\substack{+2\\-2.3}$ & N     & 3     & 2.6   & 7.6   & 1.75 \\
Kepler-334 & Tr    & Y     & 0.253 & $0.330\substack{+0.035\\-0.036}$ & $0.752\substack{+0.048\\-0.044}$ & $55.1\substack{+23.3\\-16.3}$ & Y     & 3     & 1.7   & 3.5   & 1.70 \\
    \hline \hline
    \end{tabular}
\end{table*}

\begin{table*}
\small
\centering
\ContinuedFloat
\label{}
\caption[]{continued}
\begin{tabular}{llllllllllll}
    \hline \hline
    Host name & Dis.$^{a}$ & $F_{1}$ & $\frac{\chi^2}{dof}$ & m & b     & \makecell{Period\\(days)} & $F_{2}$ & ON$^{b}$    & \makecell{$R_{Max}$\\($R_{\oplus}$)} & \makecell{$M_{Max}$\\($M_{\oplus}$)} & \makecell{$P_{tr}$\\(\%)} \\
    \hline
Kepler-102 & Tr    & Y     & 0.898 & $0.179\substack{+0.014\\-0.013}$ & $0.688\substack{+0.033\\-0.035}$ & $38.2\substack{+10.4\\-7.9}$ & Y     & 5     & 0.8   & 0.9   & 1.66 \\
Kepler-92 & Tr    & Y     & 0.032 & $0.278\substack{+0.010\\-0.010}$ & $1.141\substack{+0.012\\-0.014}$ & $94.3\substack{+10\\-8.9}$ & Y     & 3     & 2.4   & 6.6   & 1.64 \\
Kepler-445 & Tr    & N     & 0.013 & $0.219\substack{+0.005\\-0.004}$ & $0.473\substack{+0.006\\-0.007}$ & $13.4\substack{+0.7\\-0.6}$ & N     & 3 H   & 1.4   & 2.5   & 1.61 \\
Kepler-127 & Tr    & N     & 0.604 & $0.266\substack{+0.053\\-0.038}$ & $1.169\substack{+0.054\\-0.067}$ & $92.8\substack{+58.5\\-31.6}$ & N     & 3     & 2.2   & 5.6   & 1.60 \\
TOI-270 & Tr    & N     & 0.427 & $0.265\substack{+0.033\\-0.034}$ & $0.515\substack{+0.047\\-0.045}$ & $20.4\substack{+8.1\\-5.8}$ & N     & 3 H   & 2.0   & 4.7   & 1.59 \\
Kepler-54 & Tr    & Y     & 0.451 & $0.211\substack{+0.032\\-0.024}$ & $0.892\substack{+0.031\\-0.043}$ & $33.4\substack{+11.3\\-7.8}$ & Y     & 3     & 1.6   & 3.2   & 1.58 \\
Kepler-164 & Tr    & Y     & 0.257 & $0.380\substack{+0.039\\-0.042}$ & $0.689\substack{+0.051\\-0.053}$ & $67.4\substack{+31.5\\-22.8}$ & N     & 3 C   & 2.7   & 8.2   & 1.56 \\
Kepler-224 & Tr    & Y     & 0.215 & $0.261\substack{+0.015\\-0.016}$ & $0.508\substack{+0.029\\-0.028}$ & $35.6\substack{+8.1\\-6.8}$ & Y     & 4     & 2.3   & 6.1   & 1.56 \\
Kepler-257 & Tr    & Y     & 0.353 & $0.506\substack{+0.052\\-0.060}$ & $0.352\substack{+0.074\\-0.075}$ & $73.9\substack{+52\\-33}$ & Y     & 3     & 6.2   & 37.1  & 1.54 \\
Kepler-244 & Tr    & Y     & 0.083 & $0.334\substack{+0.021\\-0.020}$ & $0.642\substack{+0.026\\-0.031}$ & $44\substack{+9.9\\-8.4}$ & Y     & 3     & 1.2   & 1.9   & 1.53 \\
Kepler-295 & Tr    & Y     & 0.131 & $0.213\substack{+0.013\\-0.018}$ & $1.110\substack{+0.023\\-0.018}$ & $56\substack{+8.8\\-8.4}$ & Y     & 3     & 1.5   & 2.8   & 1.53 \\
Kepler-104 & Tr    & Y     & 0.026 & $0.328\substack{+0.011\\-0.009}$ & $1.054\substack{+0.013\\-0.013}$ & $109.1\substack{+12.6\\-9.9}$ & Y     & 3     & 4.3   & 19.1  & 1.49 \\
Kepler-172 & Tr    & Y     & 0.044 & $0.359\substack{+0.009\\-0.008}$ & $0.458\substack{+0.016\\-0.015}$ & $78.1\substack{+9.6\\-8.3}$ & Y     & 4     & 3.4   & 12.4  & 1.46 \\
Kepler-84 & Tr    & Y     & 0.548 & $0.253\substack{+0.015\\-0.015}$ & $0.648\substack{+0.036\\-0.035}$ & $81.9\substack{+24.3\\-18.5}$ & Y     & 5     & 2.6   & 7.6   & 1.45 \\
Kepler-247 & Tr    & Y     & 0.435 & $0.392\substack{+0.057\\-0.053}$ & $0.544\substack{+0.075\\-0.074}$ & $52.4\substack{+39.9\\-21.8}$ & Y     & 3     & 3.3   & 11.8  & 1.42 \\
Kepler-327 & Tr    & Y     & 0.520 & $0.373\substack{+0.068\\-0.047}$ & $0.385\substack{+0.065\\-0.078}$ & $31.9\substack{+27.4\\-12.7}$ & Y     & 3     & 1.6   & 3.2   & 1.41 \\
Kepler-770 & Tr    & N     & 0.726 & $0.553\substack{+0.109\\-0.100}$ & $0.140\substack{+0.128\\-0.152}$ & $62.8\substack{+116.8\\-40.7}$ & N     & 3     & 2.8   & 8.7   & 1.41 \\
Kepler-238 & Tr    & Y     & 0.822 & $0.336\substack{+0.025\\-0.025}$ & $0.386\substack{+0.061\\-0.062}$ & $116.5\substack{+62.3\\-40.6}$ & Y     & 5     & 5.3   & 27.9  & 1.40 \\
Kepler-249 & Tr    & Y     & 0.001 & $0.334\substack{+0.001\\-0.001}$ & $0.519\substack{+0.001\\-0.001}$ & $33.1\substack{+0.2\\-0.2}$ & Y     & 3 H   & 1.9   & 4.3   & 1.38 \\
Kepler-81 & Tr    & Y     & 0.313 & $0.271\substack{+0.029\\-0.030}$ & $0.786\substack{+0.039\\-0.040}$ & $39.7\substack{+13.3\\-10.2}$ & Y     & 3     & 1.4   & 2.5   & 1.36 \\
Kepler-83 & Tr    & Y     & 0.079 & $0.296\substack{+0.023\\-0.014}$ & $0.706\substack{+0.020\\-0.031}$ & $39.3\substack{+9.1\\-6.2}$ & Y     & 3     & 2.8   & 8.7   & 1.36 \\
K2-16 & Tr    & N     & 0.069 & $0.423\substack{+0.021\\-0.020}$ & $0.444\substack{+0.025\\-0.027}$ & $51.7\substack{+11.5\\-9.5}$ & N     & 3     & 2.2   & 5.6   & 1.28 \\
K2-58 & Tr    & N     & 0.105 & $0.474\substack{+0.028\\-0.026}$ & $0.396\substack{+0.048\\-0.040}$ & $65.5\substack{+22.9\\-15.7}$ & N     & 3     & 2.2   & 5.6   & 1.28 \\
Kepler-106 & Tr    & Y     & 0.331 & $0.280\substack{+0.016\\-0.020}$ & $0.817\substack{+0.034\\-0.028}$ & $86.2\substack{+21.4\\-18.9}$ & Y     & 4     & 1.3   & 2.2   & 1.26 \\
Kepler-122 & Tr    & Y     & 0.907 & $0.246\substack{+0.021\\-0.021}$ & $0.813\substack{+0.047\\-0.050}$ & $110.3\substack{+47\\-32.8}$ & Y     & 5     & 2.1   & 5.2   & 1.25 \\
Kepler-299 & Tr    & Y     & 0.066 & $0.368\substack{+0.010\\-0.011}$ & $0.465\substack{+0.022\\-0.019}$ & $86.7\substack{+13.2\\-11.5}$ & Y     & 4     & 2.3   & 6.1   & 1.25 \\
KOI-94 & Tr    & N     & 0.185 & $0.384\substack{+0.023\\-0.020}$ & $0.594\substack{+0.035\\-0.041}$ & $134.8\substack{+45.5\\-32.6}$ & N     & 4     & 4.2   & 18.3  & 1.25 \\
K2-32 & Tr    & N     & 0.960 & $0.297\substack{+0.035\\-0.032}$ & $0.658\substack{+0.061\\-0.062}$ & $69.8\substack{+40.7\\-24.6}$ & N     & 4     & 1.5   & 2.8   & 1.23 \\
Kepler-222 & Tr    & Y     & 0.037 & $0.428\substack{+0.023\\-0.014}$ & $0.588\substack{+0.019\\-0.030}$ & $74.6\substack{+16.9\\-11.6}$ & Y     & 3     & 4.7   & 22.4  & 1.21 \\
Kepler-282 & Tr    & Y     & 0.501 & $0.230\substack{+0.019\\-0.021}$ & $0.940\substack{+0.042\\-0.037}$ & $72.7\substack{+22.3\\-17.6}$ & Y     & 4     & 1.7   & 3.5   & 1.18 \\
Kepler-53 & Tr    & Y     & 0.065 & $0.298\substack{+0.017\\-0.015}$ & $0.983\substack{+0.019\\-0.022}$ & $75.5\substack{+13.1\\-11}$ & Y     & 3     & 3.5   & 13.1  & 1.17 \\
Kepler-332 & Tr    & Y     & 0.003 & $0.326\substack{+0.004\\-0.004}$ & $0.881\substack{+0.005\\-0.005}$ & $72.2\substack{+2.7\\-2.6}$ & Y     & 3     & 1.4   & 2.5   & 1.07 \\
K2-233 & Tr    & N     & 0.141 & $0.502\substack{+0.050\\-0.031}$ & $0.373\substack{+0.045\\-0.057}$ & $75.8\substack{+42.9\\-22}$ & N     & 3     & 2.4   & 6.6   & 1.06 \\
Kepler-215 & Tr    & Y     & 0.893 & $0.290\substack{+0.036\\-0.035}$ & $0.933\substack{+0.062\\-0.070}$ & $123.9\substack{+75.2\\-47.5}$ & Y     & 4     & 2.0   & 4.7   & 1.05 \\
Kepler-245 & Tr    & Y     & 0.073 & $0.354\substack{+0.011\\-0.009}$ & $0.515\substack{+0.019\\-0.022}$ & $85.3\substack{+13.5\\-11}$ & Y     & 4     & 3.2   & 11.1  & 1.05 \\
Kepler-325 & Tr    & Y     & 0.027 & $0.467\substack{+0.019\\-0.014}$ & $0.650\substack{+0.019\\-0.024}$ & $112.4\substack{+21.6\\-16.2}$ & Y     & 3     & 3.6   & 13.8  & 1.05 \\
K2-72 & Tr    & N     & 0.980 & $0.219\substack{+0.027\\-0.032}$ & $0.722\substack{+0.057\\-0.056}$ & $39.6\substack{+18\\-13.7}$ & N     & 4 H   & 1.5   & 2.8   & 1.03 \\
Kepler-79 & Tr    & Y     & 0.631 & $0.260\substack{+0.024\\-0.027}$ & $1.155\substack{+0.050\\-0.042}$ & $157.1\substack{+62.5\\-45.6}$ & Y     & 4     & 4.1   & 17.5  & 1.00 \\
Kepler-1388 & Tr    & Y     & 0.477 & $0.276\substack{+0.013\\-0.017}$ & $0.767\substack{+0.025\\-0.024}$ & $74.2\substack{+14.6\\-14.0}$ & Y     & 4 C,H & 2.9   & 9.3   & 0.97 \\
Kepler-154 & Tr    & Y     & 0.968 & $0.290\substack{+0.024\\-0.024}$ & $0.664\substack{+0.060\\-0.059}$ & $130.6\substack{+68\\-44.2}$ & N     & 5     & 2.9   & 9.3   & 0.96 \\
Kepler-20 & Tr    & Y     & 0.421 & $0.261\substack{+0.011\\-0.011}$ & $0.532\substack{+0.034\\-0.033}$ & $125.8\substack{+33.1\\-25.5}$ & Y     & 6     & 1.6   & 3.2   & 0.92 \\
Kepler-52 & Tr    & Y     & 0.039 & $0.333\substack{+0.016\\-0.014}$ & $0.891\substack{+0.018\\-0.019}$ & $77.8\substack{+12.9\\-10.4}$ & Y     & 3 H   & 2.4   & 6.6   & 0.89 \\
Kepler-176 & Tr    & Y     & 0.170 & $0.324\substack{+0.018\\-0.015}$ & $0.754\substack{+0.029\\-0.032}$ & $112\substack{+28.9\\-21.8}$ & Y     & 4     & 1.8   & 3.9   & 0.87 \\
Kepler-171 & Tr    & Y     & 0.210 & $0.487\substack{+0.044\\-0.046}$ & $0.602\substack{+0.058\\-0.053}$ & $115.7\substack{+63.2\\-41}$ & Y     & 3     & 2.5   & 7.1   & 0.86 \\
HD 20781 & RV    & N     & 0.261 & $0.393\substack{+0.025\\-0.024}$ & $0.728\substack{+0.047\\-0.045}$ & $200.4\substack{+79.7\\-55.4}$ & N     & 4 H   & 6.3   & 38.0  & 0.85 \\
Kepler-31 & Tr    & Y     & 0.001 & $0.312\substack{+0.001\\-0.001}$ & $1.319\substack{+0.002\\-0.001}$ & $179.5\substack{+1.9\\-1.8}$ & Y     & 3     & 4.7   & 22.4  & 0.85 \\
Kepler-229 & Tr    & Y     & 0.001 & $0.409\substack{+0.001\\-0.001}$ & $0.796\substack{+0.001\\-0.001}$ & $105.8\substack{+0.4\\-0.5}$ & Y     & 3     & 4.5   & 20.7  & 0.84 \\
Kepler-331 & Tr    & Y     & 0.101 & $0.289\substack{+0.019\\-0.020}$ & $0.935\substack{+0.024\\-0.029}$ & $63.5\substack{+13.1\\-11.7}$ & Y     & 3 H   & 1.9   & 4.3   & 0.83 \\
Kepler-357 & Tr    & Y     & 0.072 & $0.440\substack{+0.021\\-0.023}$ & $0.803\substack{+0.030\\-0.028}$ & $132.9\substack{+32\\-26.9}$ & Y     & 3     & 3.9   & 16.0  & 0.83 \\
Kepler-288 & Tr    & Y     & 0.024 & $0.485\substack{+0.018\\-0.014}$ & $0.790\substack{+0.018\\-0.022}$ & $175.7\substack{+31.3\\-23.6}$ & Y     & 3     & 3.5   & 13.1  & 0.80 \\
    \hline \hline
    \end{tabular}
\end{table*}

\begin{table*}
\small
\centering
\ContinuedFloat
\label{}
\caption[]{continued}
\begin{tabular}{llllllllllll}
    \hline \hline
    Host name & Dis.$^{a}$ & $F_{1}$ & $\frac{\chi^2}{dof}$ & m & b     & \makecell{Period\\(days)} & $F_{2}$ & ON$^{b}$    & \makecell{$R_{Max}$\\($R_{\oplus}$)} & \makecell{$M_{Max}$\\($M_{\oplus}$)} & \makecell{$P_{tr}$\\(\%)} \\
    \hline
Kepler-55 & Tr    & Y     & 0.900 & $0.335\substack{+0.025\\-0.025}$ & $0.349\substack{+0.066\\-0.063}$ & $105.6\substack{+58.3\\-37}$ & Y     & 5 H   & 2.8   & 8.7   & 0.77 \\
K2-3  & Tr    & N     & 0.850 & $0.322\substack{+0.056\\-0.058}$ & $1.023\substack{+0.071\\-0.076}$ & $97.4\substack{+71.8\\-42.8}$ & N     & 3 H   & 1.8   & 3.9   & 0.74 \\
K2-155 & Tr    & N     & 0.516 & $0.404\substack{+0.058\\-0.057}$ & $0.782\substack{+0.073\\-0.071}$ & $98.4\substack{+75.2\\-42.1}$ & N     & 3 H   & 2.4   & 6.6   & 0.74 \\
Kepler-399 & Tr    & Y     & 0.276 & $0.304\substack{+0.035\\-0.031}$ & $1.146\substack{+0.041\\-0.043}$ & $114\substack{+44.9\\-30.4}$ & Y     & 3     & 1.6   & 3.2   & 0.70 \\
TRAPPIST-1 & Tr    & N     & 0.574 & $0.181\substack{+0.007\\-0.006}$ & $0.213\substack{+0.024\\-0.023}$ & $30\substack{+5.2\\-4.3}$ & N     & 7     & 0.8   & 0.9   & 0.68 \\
Kepler-19 & Tr    & N     & 0.659 & $0.415\substack{+0.061\\-0.067}$ & $0.991\substack{+0.083\\-0.081}$ & $172\substack{+145.6\\-81.9}$ & N     & 3     & 4.6   & 21.5  & 0.67 \\
Kepler-235 & Tr    & Y     & 0.032 & $0.383\substack{+0.008\\-0.007}$ & $0.522\substack{+0.016\\-0.016}$ & $113.3\substack{+13.2\\-11}$ & Y     & 4 H   & 2.5   & 7.1   & 0.66 \\
Kepler-130 & Tr    & Y     & 0.002 & $0.507\substack{+0.004\\-0.005}$ & $0.929\substack{+0.006\\-0.006}$ & $282.4\substack{+12.5\\-12.9}$ & Y     & 3     & 2.2   & 5.6   & 0.62 \\
Kepler-166 & Tr    & Y     & 0.019 & $0.672\substack{+0.023\\-0.018}$ & $0.197\substack{+0.026\\-0.029}$ & $163.1\substack{+39.9\\-28}$ & Y     & 3 H   & 3.5   & 13.1  & 0.61 \\
Kepler-296 & Tr    & Y     & 0.033 & $0.256\substack{+0.004\\-0.004}$ & $0.773\substack{+0.008\\-0.010}$ & $113.6\substack{+8.2\\-7.2}$ & Y     & 5 H   & 2.1   & 5.2   & 0.61 \\
Kepler-289 & Tr    & N     & 0.001 & $0.281\substack{+0.001\\-0.001}$ & $1.539\substack{+0.001\\-0.001}$ & $240.4\substack{+1.9\\-1.7}$ & N     & 3     & 3.5   & 13.1  & 0.59 \\
Kepler-51 & Tr    & Y     & 0.825 & $0.231\substack{+0.046\\-0.040}$ & $1.668\substack{+0.050\\-0.058}$ & $230.1\substack{+124.4\\-77.4}$ & Y     & 3     & 10.7  & 100.0 & 0.58 \\
Kepler-251 & Tr    & Y     & 0.891 & $0.423\substack{+0.046\\-0.046}$ & $0.712\substack{+0.100\\-0.093}$ & $252.6\substack{+232.8\\-119.5}$ & Y     & 4 H   & 3.5   & 13.1  & 0.53 \\
HD 20794 & RV    & N     & 0.702 & $0.305\substack{+0.029\\-0.027}$ & $1.288\substack{+0.053\\-0.049}$ & $322.7\substack{+155\\-98.2}$ & N     & 4 H   & 3.3   & 11.9  & 0.52 \\
Kepler-30 & Tr    & Y     & 0.169 & $0.346\substack{+0.031\\-0.030}$ & $1.455\substack{+0.037\\-0.043}$ & $310.2\substack{+108.4\\-81.9}$ & Y     & 3 H   & 7.0   & 46.2  & 0.49 \\
Kepler-298 & Tr    & Y     & 0.966 & $0.442\substack{+0.102\\-0.075}$ & $0.981\substack{+0.103\\-0.134}$ & $202.2\substack{+317.5\\-113.7}$ & Y     & 3 H   & 3.2   & 11.1  & 0.45 \\
Kepler-401 & Tr    & Y     & 0.088 & $0.554\substack{+0.042\\-0.032}$ & $1.145\substack{+0.041\\-0.054}$ & $640.4\substack{+299.4\\-188.2}$ & Y     & 3 H   & 3.1   & 10.5  & 0.40 \\
Kepler-603 & Tr    & Y     & 0.754 & $0.656\substack{+0.137\\-0.111}$ & $0.753\substack{+0.155\\-0.157}$ & $527.1\substack{+1408.5\\-356.9}$ & Y     & 3 H   & 4.0   & 16.7  & 0.36 \\
GJ 3293 & RV    & N     & 0.871 & $0.311\substack{+0.033\\-0.036}$ & $1.130\substack{+0.061\\-0.072}$ & $237.1\substack{+132\\-92.8}$ & N     & 4     & 6.9   & 45.5  & 0.33 \\
HD 69830 & RV    & N     & 0.607 & $0.679\substack{+0.110\\-0.110}$ & $0.892\substack{+0.110\\-0.133}$ & $850.5\substack{+1496\\-557.7}$ & N     & 3     & 10.7  & 99.8  & 0.25 \\
Kepler-174 & Tr    & Y     & 0.824 & $0.625\substack{+0.096\\-0.114}$ & $1.101\substack{+0.136\\-0.129}$ & $944.4\substack{+1567.8\\-625.4}$ & Y     & 3     & 3.1   & 10.5  & 0.18 \\
tau Cet & RV    & N     & 0.481 & $0.499\substack{+0.040\\-0.041}$ & $1.253\substack{+0.078\\-0.071}$ & $1778.9\substack{+1300.9\\-739.7}$ & N     & 4     & 3.5   & 13.1  & 0.14 \\
HD 10180 & RV    & Y     & 0.617 & $0.513\substack{+0.026\\-0.024}$ & $0.699\substack{+0.080\\-0.078}$ & $5965.2\substack{+4366\\-2405.9}$ & Y     & 6     & 12.1  & 209.7 & 0.08 \\
GJ 676 A & RV    & N     & 0.669 & $1.136\substack{+0.100\\-0.110}$ & $0.543\substack{+0.186\\-0.194}$ & $121691.5\substack{+347768.7\\-93338.3}$ & N     & 4     & -     & -     & 0.01 \\
HR 8799 & Im    & Y     & 0.114 & $0.317\substack{+0.092\\-0.086}$ & $4.273\substack{+0.165\\-0.157}$ & $348901.9\substack{+841068.7\\-238461.9}$ & Y     & 4     & -     & -     & 0.01 \\
HD 31527 & RV    & N     & 0.764 & $0.609\substack{+0.109\\-0.116}$ & $1.186\substack{+0.147\\-0.136}$ & $1029.3\substack{+2046.7\\-692.1}$ & N     & 3     & 7.8   & 55.9  & - \\
HD 136352 & RV    & N     & 0.989 & $0.484\substack{+0.094\\-0.099}$ & $1.028\substack{+0.121\\-0.127}$ & $302.2\substack{+462.8\\-188.2}$ & N     & 3 H   & 5.4   & 28.6  & - \\
Kepler-402 & Tr    & Y     & 0.952 & $0.150\substack{+0.017\\-0.017}$ & $0.623\substack{+0.032\\-0.030}$ & $16.7\substack{+4.3\\-3.3}$ & N     & 4 C   & 1.6   & 3.2   & - \\
    \hline \hline
    \end{tabular}
    \medskip
  \tabnote{$^{a}$Discovery method of the system: 'Tr', 'RV', and 'Im' represent transit, radial velocity, and imaging, respectively.}
  \tabnote{$^b$Orbital numbers (ON) followed by 'H' indicate the predicted planets within the HZ, and ON followed by 'C' indicate that the corresponding orbital periods have been flagged as "Planetary Candidate" in NASA Exoplanet Archive.}
\end{table*}

\begin{landscape}
\begin{table}
\small
  \centering
  \caption{Systems with interpolated and extrapolated planet predictions. Columns 1 and 2 present the host star name and discovery method (Dis.). Column 3 reports a flag that defines whether the system has already been analysed by \citetalias{2015MNRAS.448.3608B} (or \citetalias{2013MNRAS.435.1126B}) (Y) or not (N). Columns 4 and 5 present $\chi^2/dof$ before and after interpolation. Columns 6 to 10, respectively, list the $\gamma$, $\Delta\gamma$, slope (m), intercept (b), and predicted orbital period. Column 11 reports whether the predicted period values in this paper and \citetalias{2015MNRAS.448.3608B} (or \citetalias{2013MNRAS.435.1126B}) are consistent within error (Y) or not (N). Columns 12 to 15, respectively, list the orbital number (ON), estimated maximum radius ($R_{Max}$), and maximum mass ($M_{Max}$) in the Earth radius and mass unit, and the transit probability ($P_{tr}$).}
\begin{tabular}{lllllllllllllll}
    \hline \hline
    Host name & Dis.$^{a}$  & $F_{1}$ & $(\frac{\chi^2}{dof})_{i}$ & $(\frac{\chi^2}{dof})_{j}$ & $\gamma$ & $\Delta\gamma$$^{b}$ & \multicolumn{1}{l}{m} & \multicolumn{1}{l}{b} & \makecell{Period\\(days)} & $F_{2}$ & ON$^{c}$    & \makecell{$R_{Max}$\\($R_{\oplus}$)} & \makecell{$M_{Max}$\\($M_{\oplus}$)} & \makecell{$P_{tr}$\\(\%)} \\
    \hline
K2-183 & Tr    & N     & 7.718 & 0.024 & 107.1 & 3.1   & \multicolumn{1}{l}{$0.338\substack{+0.003\\-0.003}$} & \multicolumn{1}{l}{$-0.330\substack{+0.010\\-0.010}$} & $1\substack{+0.1\\-0.1}$ & N     & 1     & 1.3   & 2.2   & 22.31 \\
      &       &       &       &       &       &       &       &       & $2.1\substack{+0.1\\-0.1}$ & N     & 2     & 1.6   & 3.2   & 14.87 \\
      &       &       &       &       &       &       &       &       & $4.9\substack{+0.2\\-0.2}$ & N     & 3     & 2.0   & 4.7   & 8.11 \\
      &       &       &       &       &       &       &       &       & $49.8\substack{+3.6\\-3.2}$ & N     & 6 E   & 3.5   & 13.1  & 1.72 \\
Kepler-342 & Tr    & Y     & 10.136 & 0.203 & 16.3  & 0.7   & \multicolumn{1}{l}{$0.236\substack{+0.005\\-0.005}$} & \multicolumn{1}{l}{$0.212\substack{+0.019\\-0.017}$} & $2.7\substack{+0.2\\-0.1}$ & Y     & 1     & 1.0   & 1.3   & 16.91 \\
      &       &       &       &       &       &       &       &       & $4.7\substack{+0.3\\-0.3}$ & Y     & 2     & 1.2   & 1.9   & 11.27 \\
      &       &       &       &       &       &       &       &       & $8.5\substack{+0.7\\-0.6}$ & N     & 3     & 1.3   & 2.2   & 0.80 \\
\hline \hline
    \end{tabular}
  \label{table4}
\end{table}
\end{landscape}

\begin{landscape}
\begin{table}
\small
\centering
\ContinuedFloat
\label{}
\caption[]{continued}
\begin{tabular}{lllllllllllllll}
    \hline \hline
    Host name & Dis.$^{a}$  & $F_{1}$ & $(\frac{\chi^2}{dof})_{i}$ & $(\frac{\chi^2}{dof})_{j}$ & $\gamma$ & $\Delta\gamma$$^{b}$ & \multicolumn{1}{l}{m} & \multicolumn{1}{l}{b} & \makecell{Period\\(days)} & $F_{2}$ & ON$^{c}$    & \makecell{$R_{Max}$\\($R_{\oplus}$)} & \makecell{$M_{Max}$\\($M_{\oplus}$)} & \makecell{$P_{tr}$\\(\%)} \\
    \hline
Kepler-33 & Tr    & Y     & 3.630 & 0.217 & 5.3   & 0.2   & \multicolumn{1}{l}{$0.122\substack{+0.002\\-0.002}$} & \multicolumn{1}{l}{$0.752\substack{+0.010\\-0.010}$} & $7.5\substack{+0.2\\-0.2}$ & N     & 1     & 1.9   & 4.3   & 10.47 \\
      &       &       &       &       &       &       &       &       & $10.2\substack{+0.3\\-0.3}$ & N     & 2     & 2.0   & 4.7   & 8.37 \\
      &       &       &       &       &       &       &       &       & $17.3\substack{+0.8\\-0.8}$ & N     & 4     & 2.3   & 6.1   & 5.98 \\
      &       &       &       &       &       &       &       &       & $54\substack{+3.8\\-3.4}$ & N     & 8 E   & 3.1   & 10.5  & 2.79 \\
Kepler-198 & Tr    & N     & 3.885 & 0.073 & 26.0  & 0.3   & \multicolumn{1}{l}{$0.391\substack{+0.009\\-0.008}$} & \multicolumn{1}{l}{$0.109\substack{+0.021\\-0.023}$} & $3.2\substack{+0.2\\-0.2}$ & N     & 1     & 1.2   & 1.9   & 10.30 \\
      &       &       &       &       &       &       &       &       & $7.7\substack{+0.7\\-0.7}$ & N     & 2     & 1.6   & 3.2   & 5.77 \\
      &       &       &       &       &       &       &       &       & $115.4\substack{+19.2\\-15.7}$ & N     & 5 E   & 3.0   & 9.9   & 0.94 \\
\hline \hline
    \end{tabular}
\end{table}
\end{landscape}

\begin{landscape}
\begin{table}
\small
\centering
\ContinuedFloat
\label{}
\caption[]{continued}
\begin{tabular}{lllllllllllllll}
    \hline \hline
    Host name & Dis.$^{a}$  & $F_{1}$ & $(\frac{\chi^2}{dof})_{i}$ & $(\frac{\chi^2}{dof})_{j}$ & $\gamma$ & $\Delta\gamma$$^{b}$ & \multicolumn{1}{l}{m} & \multicolumn{1}{l}{b} & \makecell{Period\\(days)} & $F_{2}$ & ON$^{c}$    & \makecell{$R_{Max}$\\($R_{\oplus}$)} & \makecell{$M_{Max}$\\($M_{\oplus}$)} & \makecell{$P_{tr}$\\(\%)} \\
    \hline
Kepler-228 & Tr    & N     & 2.456 & 0.004 & 607.6 & 115.8 & \multicolumn{1}{l}{$0.212\substack{+0.002\\-0.002}$} & \multicolumn{1}{l}{$0.407\substack{+0.003\\-0.003}$} & $6.8\substack{+0.1\\-0.1}$ & N     & 2     & 1.9   & 4.3   & 6.64 \\
      &       &       &       &       &       &       &       &       & $18\substack{+0.4\\-0.4}$ & N     & 4 E   & 2.5   & 7.1   & 3.32 \\
Kepler-142 & Tr    & N     & 3.852 & 0.075 & 25.2  & 0.2   & \multicolumn{1}{l}{$0.325\substack{+0.007\\-0.008}$} & \multicolumn{1}{l}{$0.326\substack{+0.018\\-0.018}$} & $9.5\substack{+0.8\\-0.7}$ & N     & 2     & 1.5   & 2.8   & 6.49 \\
      &       &       &       &       &       &       &       &       & $19.8\substack{+2\\-1.8}$ & N     & 3     & 1.8   & 3.9   & 4.17 \\
      &       &       &       &       &       &       &       &       & $89.7\substack{+12.2\\-11}$ & N     & 5 E   & 2.6   & 7.6   & 1.50 \\
Kepler-49 & Tr    & Y     & 2.424 & 0.085 & 27.5  & 13.1  & \multicolumn{1}{l}{$0.213\substack{+0.005\\-0.005}$} & \multicolumn{1}{l}{$0.415\substack{+0.013\\-0.012}$} & $4.2\substack{+0.2\\-0.2}$ & Y     & 1     & 1.1   & 1.6   & 6.13 \\
      &       &       &       &       &       &       &       &       & $30\substack{+2.9\\-2.5}$ & Y     & 5 E   & 1.8   & 3.9   & 1.72 \\
\hline \hline
    \end{tabular}
\end{table}
\end{landscape}

\begin{landscape}
\begin{table}
\small
\centering
\ContinuedFloat
\label{}
\caption[]{continued}
\begin{tabular}{lllllllllllllll}
    \hline \hline
    Host name & Dis.$^{a}$  & $F_{1}$ & $(\frac{\chi^2}{dof})_{i}$ & $(\frac{\chi^2}{dof})_{j}$ & $\gamma$ & $\Delta\gamma$$^{b}$ & \multicolumn{1}{l}{m} & \multicolumn{1}{l}{b} & \makecell{Period\\(days)} & $F_{2}$ & ON$^{c}$    & \makecell{$R_{Max}$\\($R_{\oplus}$)} & \makecell{$M_{Max}$\\($M_{\oplus}$)} & \makecell{$P_{tr}$\\(\%)} \\
    \hline
Kepler-218 & Tr    & N     & 3.612 & 0.133 & 26.1  & 0.6   & \multicolumn{1}{l}{$0.518\substack{+0.024\\-0.021}$} & \multicolumn{1}{l}{$0.570\substack{+0.039\\-0.049}$} & $12.3\substack{+1.9\\-1.8}$ & N     & 1     & 1.5   & 2.8   & 4.64 \\
      &       &       &       &       &       &       &       &       & $437.8\substack{+160.6\\-114.6}$ & N     & 4 E,H & 3.6   & 13.8  & 0.43 \\
K2-37 & Tr    & N     & 2.611 & 0.012 & 222.9 & 44.2  & \multicolumn{1}{l}{$0.168\substack{+0.002\\-0.002}$} & \multicolumn{1}{l}{$0.645\substack{+0.004\\-0.004}$} & $9.5\substack{+0.2\\-0.2}$ & N     & 2     & 2.0   & 4.7   & 4.60 \\
      &       &       &       &       &       &       &       &       & $20.7\substack{+0.6\\-0.6}$ & N     & 4 E   & 2.4   & 6.6   & 2.79 \\
Kepler-354 & Tr    & N     & 5.502 & 0.007 & 420.1 & 30.1  & \multicolumn{1}{l}{$0.162\substack{+0.001\\-0.001}$} & \multicolumn{1}{l}{$0.739\substack{+0.003\\-0.003}$} & $7.9\substack{+0.1\\-0.1}$ & N     & 1     & 0.9   & 1.1   & 4.60 \\
      &       &       &       &       &       &       &       &       & $11.5\substack{+0.1\\-0.1}$ & N     & 2     & 1.0   & 1.3   & 3.42 \\
      &       &       &       &       &       &       &       &       & $35.4\substack{+0.7\\-0.7}$ & N     & 5 E   & 1.4   & 2.5   & 1.71 \\
\hline \hline
    \end{tabular}
\end{table}
\end{landscape}

\begin{landscape}
\begin{table}
\small
\centering
\ContinuedFloat
\label{}
\caption[]{continued}
\begin{tabular}{lllllllllllllll}
    \hline \hline
    Host name & Dis.$^{a}$  & $F_{1}$ & $(\frac{\chi^2}{dof})_{i}$ & $(\frac{\chi^2}{dof})_{j}$ & $\gamma$ & $\Delta\gamma$$^{b}$ & \multicolumn{1}{l}{m} & \multicolumn{1}{l}{b} & \makecell{Period\\(days)} & $F_{2}$ & ON$^{c}$    & \makecell{$R_{Max}$\\($R_{\oplus}$)} & \makecell{$M_{Max}$\\($M_{\oplus}$)} & \makecell{$P_{tr}$\\(\%)} \\
    \hline
HIP 41378 & Tr    & N     & 2.078 & 0.804 & 0.3   & 0.1   & \multicolumn{1}{l}{$0.150\substack{+0.003\\-0.003}$} & \multicolumn{1}{l}{$-0.723\substack{+0.009\\-0.007}$} & $22.1\substack{+0.6\\-0.5}$ & N     & 1     & 2.3   & 6.1   & 3.80 \\
      &       &       &       &       &       &       &       &       & $44.8\substack{+2\\-1.7}$ & N     & 3     & 2.8   & 8.7   & 2.37 \\
      &       &       &       &       &       &       &       &       & $82.8\substack{+5.4\\-4.8}$ & N     & 5     & 3.2   & 11.1  & 1.57 \\
      &       &       &       &       &       &       &       &       & $266.5\substack{+21.4\\-18.8}$ & N     & 8     & 4.3   & 19.1  & 0.72 \\
      &       &       &       &       &       &       &       &       & $301.4\substack{+33.3\\-29}$ & N     & 9     & 4.5   & 20.7  & 0.67 \\
      &       &       &       &       &       &       &       &       & $701.2\substack{+78.3\\-67.8}$ & N     & 11 E,H & 5.5   & 29.8  & 0.38 \\
Kepler-46 & Tr    & N     & 5.195 & 0.003 & 861.2 & 93.4  & \multicolumn{1}{l}{$0.231\substack{+0.001\\-0.001}$} & \multicolumn{1}{l}{$0.831\substack{+0.003\\-0.002}$} & $11.6\substack{+0.1\\-0.1}$ & N     & 1     & 1.9   & 4.3   & 3.75 \\
\hline \hline
    \end{tabular}
\end{table}
\end{landscape}

\begin{landscape}
\begin{table}
\small
\centering
\ContinuedFloat
\label{}
\caption[]{continued}
\begin{tabular}{lllllllllllllll}
    \hline \hline
    Host name & Dis.$^{a}$  & $F_{1}$ & $(\frac{\chi^2}{dof})_{i}$ & $(\frac{\chi^2}{dof})_{j}$ & $\gamma$ & $\Delta\gamma$$^{b}$ & \multicolumn{1}{l}{m} & \multicolumn{1}{l}{b} & \makecell{Period\\(days)} & $F_{2}$ & ON$^{c}$    & \makecell{$R_{Max}$\\($R_{\oplus}$)} & \makecell{$M_{Max}$\\($M_{\oplus}$)} & \makecell{$P_{tr}$\\(\%)} \\
    \hline
Kepler-37 & Tr    & Y     & 1.529 & 0.097 & 4.9   & 1.3   & \multicolumn{1}{l}{$0.095\substack{+0.002\\-0.002}$} & \multicolumn{1}{l}{$1.129\substack{+0.005\\-0.006}$} & $16.9\substack{+0.3\\-0.3}$ & Y     & 1     & 0.4   & 0.3   & 3.03 \\
      &       &       &       &       &       &       &       &       & $25.6\substack{+0.6\\-0.6}$ & Y     & 3     & 0.4   & 0.3   & 2.27 \\
      &       &       &       &       &       &       &       &       & $32.3\substack{+0.9\\-0.9}$ & Y     & 4     & 0.4   & 0.3   & 1.91 \\
      &       &       &       &       &       &       &       &       & $62.7\substack{+2.4\\-2.3}$ & Y     & 7 E   & 0.5   & 0.4   & 1.25 \\
Kepler-267 & Tr    & N     & 2.195 & 0.001 & 5133.5 & 327.8 & \multicolumn{1}{l}{$0.309\substack{+0.001\\-0.001}$} & \multicolumn{1}{l}{$0.527\substack{+0.001\\-0.001}$} & $14\substack{+0.1\\-0.1}$ & N     & 2     & 1.9   & 4.3   & 2.75 \\
      &       &       &       &       &       &       &       &       & $58.1\substack{+0.5\\-0.5}$ & N     & 4 E   & 2.7   & 8.2   & 1.07 \\
GJ 667 C & RV    & N     & 3.514 & 0.306 & 2.6   & 0.6   & \multicolumn{1}{l}{$0.192\substack{+0.005\\-0.004}$} & \multicolumn{1}{l}{$0.856\substack{+0.017\\-0.019}$} & $11.9\substack{+0.6\\-0.6}$ & N     & 1     & 0.2   & 0.1   & 2.75 \\
\hline \hline
    \end{tabular}
\end{table}
\end{landscape}

\begin{landscape}
\begin{table}
\small
\centering
\ContinuedFloat
\label{}
\caption[]{continued}
\begin{tabular}{lllllllllllllll}
    \hline \hline
    Host name & Dis.$^{a}$  & $F_{1}$ & $(\frac{\chi^2}{dof})_{i}$ & $(\frac{\chi^2}{dof})_{j}$ & $\gamma$ & $\Delta\gamma$$^{b}$ & \multicolumn{1}{l}{m} & \multicolumn{1}{l}{b} & \makecell{Period\\(days)} & $F_{2}$ & ON$^{c}$    & \makecell{$R_{Max}$\\($R_{\oplus}$)} & \makecell{$M_{Max}$\\($M_{\oplus}$)} & \makecell{$P_{tr}$\\(\%)} \\
    \hline
WASP-47 & Tr    & N     & 6.156 & 0.410 & 7.0   & 0.7   & \multicolumn{1}{l}{$0.568\substack{+0.022\\-0.022}$} & \multicolumn{1}{l}{$-0.068\substack{+0.065\\-0.066}$} & $43.5\substack{+15.3\\-11.5}$ & N     & 3     & 4.9   & 24.2  & 2.14 \\
      &       &       &       &       &       &       &       &       & $161.8\substack{+68.1\\-48.3}$ & N     & 4     & 6.8   & 43.9  & 0.89 \\
      &       &       &       &       &       &       &       &       & $2179.7\substack{+1266.1\\-808.7}$ & N     & 6 E   & 13.1  & -     & 0.16 \\
Kepler-178 & Tr    & N     & 2.330 & 0.001 & 3627.8 & 1221.9 & \multicolumn{1}{l}{$0.335\substack{+0.001\\-0.001}$} & \multicolumn{1}{l}{$0.980\substack{+0.002\\-0.002}$} & $44.6\substack{+0.4\\-0.4}$ & N     & 2     & 3.3   & 11.8  & 2.05 \\
      &       &       &       &       &       &       &       &       & $208.7\substack{+2.9\\-3.1}$ & N     & 4 E   & 4.8   & 23.3  & 0.75 \\
GJ 163 & RV    & N     & 4.847 & 0.005 & 510.2 & 47.4  & \multicolumn{1}{l}{$0.462\substack{+0.003\\-0.003}$} & \multicolumn{1}{l}{$0.941\substack{+0.007\\-0.006}$} & $72.9\substack{+2.1\\-2}$ & N     & 2 H   & 1.4   & 2.5   & 2.05 \\
      &       &       &       &       &       &       &       &       & $214.6\substack{+7.4\\-7.3}$ & N     & 3     & 2.8   & 8.8   & 0.99 \\
\hline \hline
    \end{tabular}
\end{table}
\end{landscape}

\begin{landscape}
\begin{table}
\small
\centering
\ContinuedFloat
\label{}
\caption[]{continued}
\begin{tabular}{lllllllllllllll}
    \hline \hline
    Host name & Dis.$^{a}$  & $F_{1}$ & $(\frac{\chi^2}{dof})_{i}$ & $(\frac{\chi^2}{dof})_{j}$ & $\gamma$ & $\Delta\gamma$$^{b}$ & \multicolumn{1}{l}{m} & \multicolumn{1}{l}{b} & \makecell{Period\\(days)} & $F_{2}$ & ON$^{c}$    & \makecell{$R_{Max}$\\($R_{\oplus}$)} & \makecell{$M_{Max}$\\($M_{\oplus}$)} & \makecell{$P_{tr}$\\(\%)} \\
    \hline
HD 40307 & RV    & Y     & 1.021 & 0.211 & 3.9   & 11.5  & \multicolumn{1}{l}{$0.338\substack{+0.008\\-0.009}$} & \multicolumn{1}{l}{$0.648\substack{+0.025\\-0.025}$} & $98\substack{+14.3\\-13.4}$ & Y     & 4 H   & 1.6   & 3.1   & 0.87 \\
      &       &       &       &       &       &       &       &       & $472\substack{+88.7\\-80.5}$ & N     & 6 E   & 4.4   & 19.6  & 0.30 \\
HD 27894 & RV    & N     & 11.719 & 0.228 & 12.6  & 0.2   & \multicolumn{1}{l}{$0.414\substack{+0.010\\-0.010}$} & \multicolumn{1}{l}{$1.208\substack{+0.037\\-0.035}$} & $105.2\substack{+15.1\\-12.9}$ & N     & 2     & 4.2   & 18.3  & 0.80 \\
      &       &       &       &       &       &       &       &       & $288.1\substack{+46.5\\-39.2}$ & N     & 3     & 8.0   & 59.2  & 0.41 \\
      &       &       &       &       &       &       &       &       & $780.9\substack{+139.9\\-116.2}$ & N     & 4     & 12.1  & 189.5 & 0.21 \\
      &       &       &       &       &       &       &       &       & $1695.9\substack{+414\\-338}$ & N     & 5     & 12.2  & 468.4 & 0.12 \\
      &       &       &       &       &       &       &       &       & $12768.7\substack{+3498.3\\-2750.3}$ & N     & 7 E   & 12.5  & 4938.6 & 0.03 \\
\hline \hline
    \end{tabular}
    \medskip
    \tabnote{$^{a}$Discovery method of the system: 'Tr', 'RV', 'OBM', and 'PT' represent transit, radial velocity, orbital brightness modulation, and pulsar timing, respectively.}
    \tabnote{$^b$$\Delta\gamma$=($\gamma_{1}$-$\gamma_{2}$)/$\gamma_{2}$, where $\gamma_{1}$ and $\gamma_{2}$ are the highest and second-highest $\gamma$ values for the system, respectively.}
    \tabnote{$^c$Orbital numbers (ON) followed by 'E' indicate the extrapolated planets, followed by 'H' indicate the predicted planets within the HZ, and followed by 'C' indicate that corresponding orbital periods have been flagged as "Planetary Candidate" in NASA Exoplanet Archive.}
\end{table}
\end{landscape}

\begin{table*}
  \centering
  \caption{Systems with detected planets since predictions made by \citetalias{2015MNRAS.448.3608B}.}
    \begin{tabular}{lccccc}
    \hline \hline
Host name & \makecell{Detected period\\(days)} & \makecell{\citetalias{2015MNRAS.448.3608B} predicted period\\(days)} & \makecell{Our predicted period\\(days)} & \makecell{\citetalias{2015MNRAS.448.3608B} error\\(\%)} & \makecell{Our error\\(\%)} \\
\hline
Kepler-1388 & 75.73 & $73.0\pm8.0$ & $74.2\substack{+14.6\\-14.0}$ & 3.6   & 2.0 \\
Kepler-150 & 637.21 & N/A   & N/A   & N/A   & N/A \\
Kepler-1542 & 7.23  & $6.9\pm0.3$ & $6.9\substack{+0.2\\-0.2}$ & 4.6   & 4.6 \\
Kepler-20 & 34.94 & $39.1\pm5.4$ & $39.0\substack{+5.9\\-5.3}$ & 11.9  & 11.6 \\
Kepler-80 & 14.65 & $14.5\pm1.3$ & $14.6\substack{+0.8\\-0.7}$ & 1.0   & 0.3 \\
Kepler-82 & 75.73 & N/A   & N/A   & N/A   & N/A \\
KOI-351 & 14.45 & $15.4\pm1.7$ & $14.3\substack{+1.0\\-0.9}$ & 6.6   & 1.0 \\
    \hline \hline
    \end{tabular}
  \label{table5}
\end{table*}

\FloatBarrier
\bibliographystyle{pasa-mnras}
\bibliography{1r_lamboo_notes}

\end{document}